\documentclass[conference]{IEEEtran}

\usepackage[utf8]{inputenc}
\usepackage[T1]{fontenc}
\usepackage{hyperref}
\usepackage{url}
\usepackage{booktabs}
\usepackage{amsfonts}
\usepackage{amsmath}
\usepackage{nicefrac}
\usepackage{microtype}
\usepackage{cleveref}
\usepackage{wrapfig}
\usepackage{float}
\usepackage{dblfloatfix}

\usepackage{nicefrac}
\usepackage{microtype}
\usepackage{lipsum}
\usepackage{graphicx}
\usepackage{pdflscape}
\usepackage{multirow}
\usepackage{booktabs}
\usepackage{xcolor}

\usepackage{amssymb}
\usepackage{pifont}
\usepackage{cleveref}
\usepackage{adjustbox}
\usepackage[numbers]{natbib}
\usepackage[ruled,vlined]{algorithm2e}

\usepackage{makecell}
\usepackage{diagbox}
\usepackage{wrapfig}
\usepackage{listings}

\usepackage{tikz}
\usepackage{tikzpagenodes}
\usetikzlibrary{positioning}
\usetikzlibrary{fit}
\usetikzlibrary{calc}

\newcommand*\circled[1]{\tikz[baseline=(char.base)]{
            \node[shape=circle,fill,inner sep=1.2pt] (char) {\textcolor{white}{#1}};}}

\newcommand{\mathdefault}[1][]{}

\definecolor{jcgreen}{HTML}{33a02c}
\definecolor{jcred}{HTML}{e31a1c}

\newcommand{\cmark}{\textcolor{jcgreen}{\ding{51}}}%
\newcommand{\xmark}{\textcolor{jcred}{\ding{55}}}%

\definecolor{darkelectricblue}{rgb}{0.33, 0.41, 0.47}

\definecolor{czgreen}{HTML}{0D8A5D}
\definecolor{czred}{HTML}{CC0000}
\definecolor{codegreen}{rgb}{0,0.6,0}
\definecolor{codegray}{rgb}{0.5,0.5,0.5}
\definecolor{codepurple}{rgb}{0.58,0,0.82}
\definecolor{backcolour}{rgb}{0.95,0.95,0.92}

\lstdefinestyle{mystyle}{
    backgroundcolor=\color{backcolour},
    commentstyle=\color{codegreen},
    keywordstyle=\color{magenta},
    numberstyle=\tiny\color{codegray},
    stringstyle=\color{codepurple},
    basicstyle=\ttfamily\footnotesize,
    breakatwhitespace=false,
    breaklines=true,
    captionpos=b,
    keepspaces=true,
    numbers=left,
    numbersep=5pt,
    showspaces=false,
    showstringspaces=false,
    showtabs=false,
    tabsize=2
}

\usepackage{cite}
\usepackage{amsmath,amssymb,amsfonts}
\usepackage{algorithmic}
\usepackage{graphicx}
\usepackage{textcomp}
\usepackage{xcolor}
\def\BibTeX{{\rm B\kern-.05em{\sc i\kern-.025em b}\kern-.08em
    T\kern-.1667em\lower.7ex\hbox{E}\kern-.125emX}}

\usepackage{placeins}
\usepackage{pdfpages}

\renewcommand{\cite}{\citep}

\definecolor{grey}{rgb}{0.5, 0.5, 0.5}
\newcommand{\disclaimer}{
	\begin{tikzpicture}[remember picture,overlay]
		\coordinate (pasouth) at (current page.south);
		\coordinate (tasouth) at (current page text area.south);
		\node[anchor=west, inner sep=0pt, xshift=-\fboxrule]
			at ({$(pasouth)!0.5!(tasouth)$} -| current page text area.west)
			{
				\color{grey}
				\fbox{\parbox{\dimexpr\textwidth-2\fboxsep-\fboxrule\relax}{
					\scriptsize
					\centering
					This work has been accepted for publication in the 2025 IEEE Conference on Secure and Trustworthy Machine Learning. The final version will be available on IEEE Xplore.
				}}
			};
	\end{tikzpicture}%
}

\AddToHook{shipout/background}{\put(0,0){\disclaimer}}

\begin{document}

\title{Locking Machine Learning Models into Hardware}

\author{
	\IEEEauthorblockN{Eleanor Clifford*}
	\IEEEauthorblockA{
		\textit{Imperial College London}\\
		eleanor.clifford@cl.cam.ac.uk
	} \and
	\IEEEauthorblockN{Adhithya Saravanan*}
	\IEEEauthorblockA{
		\textit{University of Cambridge}\\
		aps85@cam.ac.uk
	} \and
	\IEEEauthorblockN{Harry Langford*}
	\IEEEauthorblockA{
		\textit{University of Cambridge}\\
		hjel2@cam.ac.uk
	} \and
	\IEEEauthorblockN{Cheng Zhang}
	\IEEEauthorblockA{
		\textit{Imperial College London}\\
		cheng.zhang122@imperial.ac.uk
	} \and
	\IEEEauthorblockN{Yiren Zhao}
	\IEEEauthorblockA{
		\textit{Imperial College London}\\
		a.zhao@imperial.ac.uk
	} \and
	\IEEEauthorblockN{Robert Mullins}
	\IEEEauthorblockA{
		\textit{University of Cambridge}\\
		robert.mullins@cl.cam.ac.uk
	} \and
	\IEEEauthorblockN{Ilia Shumailov}
	\IEEEauthorblockA{
		\textit{Google Deepmind}\\
		iliashumailov@google.com
	} \and
	\IEEEauthorblockN{Jamie Hayes}
	\IEEEauthorblockA{
		\textit{Google Deepmind}\\
		jamhay@google.com
	}
}
\IEEEoverridecommandlockouts
\IEEEaftertitletext{\vspace{-1\baselineskip}}

\maketitle

\begingroup\renewcommand\thefootnote{*}
\footnotetext{Equal contribution}
\endgroup

\begin{abstract}

Modern machine learning (ML) models are expensive IP and business competitiveness often depends on keeping this IP confidential. This in turn restricts how these models are deployed; for example, it is unclear how to deploy a model on-device without inevitably leaking the underlying model. At the same time, confidential computing technologies such as multi-party computation or homomorphic encryption remain impractical for wide adoption. In this paper, we take a different approach and investigate the feasibility of ML-specific mechanisms that deter unauthorized model use by restricting the model to only be usable on specific hardware, making adoption on unauthorized hardware inconvenient. That way, even if IP is compromised, it cannot be trivially used without specialised hardware or major model adjustment. In a sense, we seek to enable cheap \emph{locking of machine learning models into specific hardware}. We demonstrate that \emph{locking} mechanisms are feasible by either targeting efficiency of model representations, making such models incompatible with quantization, or tying the model's operation to specific characteristics of hardware, such as the number of clock cycles for arithmetic operations. We demonstrate that locking comes with negligible overheads, while significantly restricting usability of the resultant model on unauthorized hardware.

\end{abstract}


\section{Introduction}

\begin{table*}[h]
\centering
\caption{Taxonomy of locking methods. For reverse engineering, access is given to the locked model but not the target hardware. * Similar to the cost of fine-tuning the model. ** The PUF method provides sufficient entropy to use the cheaper \emph{AES encryption} transformation.}

\adjustbox{max width=\linewidth}{

\begin{tabular}{lllll}
\toprule

   \textbf{Category}
 & \textbf{Method}
 & \textbf{Effect}
 & \textbf{Reverse-engineering cost}
 & \textbf{Overhead} \\

\midrule

 \multirow{2}{*}{Soft locking}
 & Sparsity-aware
 & Slowdown or Accuracy drop
 & Moderate*
 & Small  \\

 & Quantisation-aware
 & Slowdown or Accuracy drop
 & Moderate*
 & Small  \\

\midrule

 \multirow{3}{*}{Hard locking}

 & Clock fingerprint
 & No Accuracy
 & High
 & Moderate  \\

 & Finite Precision
 & No Accuracy
 & High
 & Moderate  \\

 & SRAM PUF
 & No Accuracy
 & Infeasible
 & Small**    \\

\bottomrule
\end{tabular}
}

\label{tab:taxonomy}
\end{table*}

The monetary expenditures associated with developing machine learning (ML) models are increasing rapidly with the advent of large generative models. Models with over a trillion parameters are now being trained on web-scale data \citep{brown2020language}.
These models have become valuable Intellectual Property (IP) assets, yet ensuring their competitive edge remains uncompromised when deployed on-device proves challenging. Competitors may reverse engineer the model's architecture and parameters, redeploying it on their software and hardware stack. Concurrently, \textit{governance} of Machine Learning models is a concern \citep{hadfield2023regulatory}. Especially in safety-critical applications, it may be necessary to limit model execution to special authenticated settings. Here, we usually rely on hardware and software combinations to prevent model use on unverified platforms, which may lead to the potential misuse of the model.

Existing ML governance and IP protection methods can be classified into two categories: namely \textit{policies} and \textit{centralised serving}. Policy-based methods focus on either access control or licensing. For example, accessing LLaMA models requires users to sign a terms of service agreement \citep{touvron2023llama}, and licenses like OpenRail \citep{ferrandis2022openrail} include usage limitations to prevent misuse of these large language models (LLMs). However, it is entirely possible for these access-controlled models to leak, as it happened to LLaMA2~\citep{llamaleak}, and malicious users may not adhere to any existing licenses, as pointed out by \citeauthor{henderson2023selfdestructing} and \citeauthor{lin2024malla}.
Another approach involves hosting ML models on centralised servers and providing standard API access to users. Companies employ this method to safeguard their IP, implementing safety filters and safeguarding prompts to ensure appropriate usage. It is worth mentioning that this centralised model serving necessitates substantial resources to maintain, as all user queries are handled by centralised computing infrastructure, unlike computations on user devices, which are typically less checked. Furthermore, these models cannot be used offline.

\begin{figure}
    \centering
    \adjustbox{width=\linewidth}{
        \includegraphics{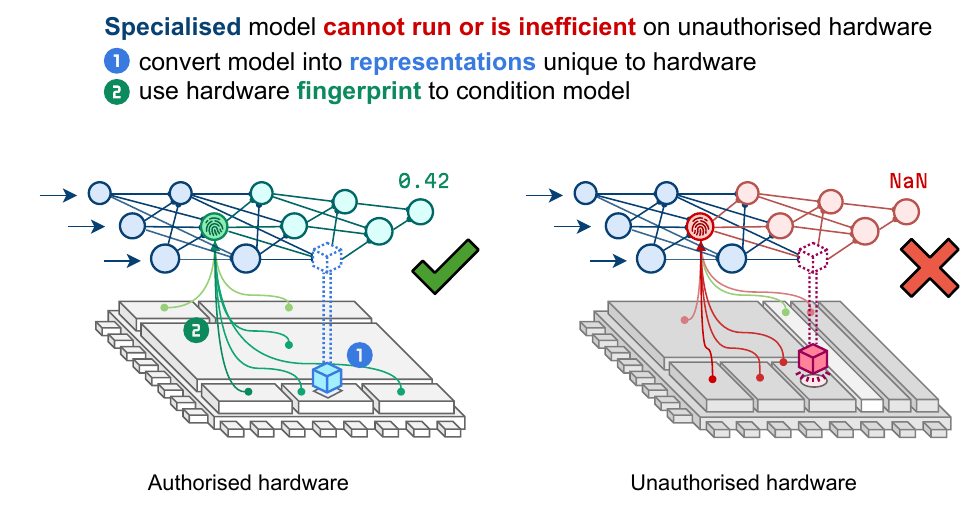}
    }
    \vspace{-5mm}
    \caption{A high-level illustration of how \textit{ML Hardware Locking} functions: the locked model resists efficient, or any, deployment by adversaries on unauthorized hardware stacks. This resistance occurs because unauthorized hardware devices inherently lack support for some hardware operation or are unable to match the hardware properties of the authorized hardware.}
    \label{fig:motivation}
\end{figure}

As both policy enforcement and centralised serving fail to address the issue of deploying whole or parts of ML models on user devices, we adopt a completely different approach from the aforementioned methods in this study, illustrated in \Cref{fig:motivation}. We explore the feasibility of mechanisms enabling \textit{ML Hardware Locking}, whereby a locked ML model resists efficient, or any, deployment by adversaries on unauthorized hardware stacks.
In such scenarios, should a model be stolen or reverse-engineered, deploying it on unauthorized platforms would be either impossible or extremely challenging.

\begin{itemize}

    \item We introduce the concept of \textit{ML Hardware Locking}, where ML models are locked to specific hardware stacks, restricting model usage on unauthenticated hardware.

    \item We demonstrate the feasibility of a range of soft and hard locking mechanisms; these methods have negligible overheads when deployed, and we quantify the difficulty of breaking these locks on unauthorized hardware platforms. An overview can be seen in~\Cref{tab:taxonomy}.

    \item We explore widely available hardware asymmetries as a foundation for AI governance and discuss its implications.

\end{itemize}

All code can be found at~\href{https://sr.ht/~ecc/MLHardwareLocking}{\nolinkurl{sr.ht/~ecc/MLHardwareLocking}}.

\section{Background}

\begin{table*}[hbt]
\centering
\caption{A comparison of the supported features of existing AI accelerators. * We consider also the TensorCore supported instruction for NVIDIA GPUs. TEE stands for the Trusted Execution Environment. ** DOJO utilised a specialised floating point arithmetic (CFP), where they have a different setup for exponent and mantissa bit widths.}
\adjustbox{max width=\linewidth}{
\begin{tabular}{lrccccccccc}
\toprule
\textbf{Device} &
& \textbf{INT4}
& \textbf{INT8}
& \textbf{FP8}
& \textbf{FP16}
& \textbf{FP32}
& \textbf{TF32}
& \textbf{BFLOAT16}
& \textbf{Sparsity} 
& \textbf{TEE}\\
\midrule
NVIDIA A100 * 
& \cite{choquette2021nvidia}
& \cmark
& \cmark
& \xmark
& \cmark
& \cmark
& \cmark
& \cmark 
& \cmark
& \xmark
\\
NVIDIA H100 * & \cite{choquette2023nvidia}
& \xmark
& \cmark
& \cmark
& \cmark
& \cmark
& \cmark
& \cmark 
& \cmark
& \cmark
\\
NVIDIA H200 * & \cite{NvidiaH200}
& \xmark
& \cmark
& \cmark
& \cmark
& \cmark
& \cmark
& \cmark 
& \cmark
& \cmark
\\
Cerebras WSE 2 & \cite{selig2022cerebras}
& \xmark
& \xmark
& \xmark
& \cmark
& \cmark
& \xmark
& \cmark
& \cmark
& \xmark
\\
Tesla DOJO & \cite{talpes2023microarchitecture}
& \xmark
& \cmark
& \cmark **
& \cmark **
& \cmark
& \xmark 
& \cmark
& \xmark
& \xmark
\\
Groq & \cite{gwennap2020groq}
& \xmark
& \cmark
& \xmark
& \cmark
& \cmark
& \xmark 
& \xmark
& \xmark
& \xmark
\\
Qualcomm AI100 & \cite{chatha2021qualcomm}
& \xmark
& \cmark
& \xmark
& \cmark
& \cmark
& \xmark
& \xmark
& \xmark
& \xmark
\\
Google TPU V4i & \cite{jouppi2021ten}
& \xmark
& \cmark
& \xmark
& \cmark
& \cmark
& \xmark
& \cmark
& \xmark
& \xmark
\\
AMD MI300 & \cite{AMD}
& \xmark
& \cmark
& \cmark
& \cmark
& \cmark
& \cmark
& \cmark
& \cmark
& \xmark
\\
AMD AIE & \cite{alok2020architecture}
& \cmark
& \cmark
& \xmark
& \cmark
& \xmark
& \xmark
& \cmark
& \xmark
& \xmark
\\
\bottomrule
\end{tabular}
}
\label{tab:related_works}
\end{table*}

\subsection{Software and Hardware}
Traditional software systems often employ measures to ensure that software executes exclusively on authorized hardware.
This is achieved by integrating hardware identifiers, or fingerprints, into the software itself, and is similar to the methods presented in this paper.
Upon execution, the software verifies the authenticity of the hardware it is running on by comparing its fingerprint to the expected value. Such fingerprints can be, for example, generated with Picasso by using web-browser agents and HTML canvases for identification~\citep{bursztein2016picasso}, or DrawnApart, which relies on GPU speed variance~\citep{Laor_2022}.

Furthermore, comprehensive verification of the software and hardware loading process is frequently implemented. This involves employing secure boot-loaders, firmware checks, hardware authentication tokens, and platform modules to ensure the integrity of the system as a whole. By combining hardware fingerprinting with a thorough verification process, these systems can effectively restrict software execution to pre-approved hardware. Note, that all of the solutions are usually used in conjunction. In this paper we build examples of such mechanisms which are specifically suited for machine learning.

\subsection{Machine Learning Deployment}

Modern machine learning deployment primarily utilises centralised serving rather than on-device inference due to challenges in data sharing and the necessity for powerful specialised hardware. In the realm of on-device inference, the prevailing approach to access and security management is through policy-based restrictions~\citep{outchakoucht2017dynamic} and safety fine tuning~\citep{qi2023fine}.

\subsection{Specialised Hardware}

AI hardware vendors have designed a wide range of chips that feature a comprehensive range of hardware intrinsic supports. This often focuses on hardware arithmetic, optimisations such as sparsity, and security support like Trusted Execution Environment (TEE).
\Cref{tab:related_works} presents a compilation of recently developed AI accelerators along with their respective hardware intrinsic supports. It details features encompassing various arithmetic supports, including \textbf{INT4}, \textbf{INT8}, \textbf{FP8}, \textbf{FP16}, \textbf{FP32}, \textbf{TF32}, and \textbf{BFLOAT16}. \Cref{tab:related_works} also considers the availability of support for sparse matrix multiplication (sparsity) and TEE in the devices listed. The device-level hardware intrinsic asymmetry, detailed in \Cref{tab:related_works}, provides substantial opportunities for its exploitation in hardware locking. While software or compiler optimisations can imitate circuit design variations, these emulations \textit{are inherently less efficient}, often by orders of magnitude, in terms of operation-per-Watt performance.

Furthermore, \Cref{tab:related_works} illustrates how the same hardware support can vary in implementation. For instance, Tesla's DOJO adopted distinctive \textbf{FP8} and \textbf{FP16} schemes, deviating from the conventional schemes, and termed them CFP. Notably, computations occur at finite precision, and the execution of the same operation often varies across hardware platforms. Consequently, even a standard \textbf{FP32} convolution operation might yield different numerical results on different hardware, as detailed by \citet{schlogl2024causes}. The differences in hardware implementations and numerical deviations described above can also serve as sources of asymmetry to explore for locking.

\subsection{Existing Security Mechanisms}

ML Hardware Locking complements existing security measures like key-based encryption, trusted execution environments (TEEs) \citep{sabt2015trusted}, multi-party computation (MPC) \citep{goldreich1998secure}, and homomorphic encryption (HEE) \citep{naehrig2011can}; and we have seen adaptations of these techniques in the field of federated learning \citep{mugunthan2019smpai, zhang2020batchcrypt, chen2020training}. While these techniques offer strong security guarantees, they often come with significant overhead in terms of performance, complexity, or specialized hardware requirements. ML Hardware Locking, in contrast, aims to provide a lightweight and potentially more accessible solution, particularly in scenarios where traditional approaches might be impractical or unavailable.

\newpage
In addition to the general approaches described above, \citet{chakraborty2020hardwareassisted} proposed a specific hardware-assisted obfuscation framework called the Hardware Protected Neural Network (HPNN). HPNN leverages a key-dependent backpropagation algorithm during the training process. This creates a model whose learned weight space is intrinsically tied to a secret key embedded within a trusted hardware device. Only authorized hardware, that is possessing the secret key, can correctly execute the model's inference. This approach aligns with our hard locking, specifically utilizing a hardware root-of-trust to bind the model's functionality to authorized devices, preventing its effective use elsewhere. The HPNN framework provides a concrete example of how hardware characteristics can be deeply integrated into the model training itself, offering a strong form of IP protection.

While HPNN focuses on modifying the training process to achieve locking, our work expands upon this by exploring a wider range of both hard locking techniques (like clock and finite precision fingerprints) and soft locking methods (like sparsity-aware and quantization-aware locking). These soft locking methods, unlike HPNN, do not fully prevent execution on unauthorized hardware, but instead aim to significantly degrade performance or accuracy, providing a different, potentially less resource-intensive level of protection. Furthermore, our proposed hard locking methods using alternative hardware fingerprints offer flexibility in terms of locking granularity (device family, model, or individual device), and can be used when a root-of-trust is not available, whereas HPNN's approach primarily targets locking to specific trusted hardware.

\section{Threat Model}

We assume an adversary who has access to the model's parameters and architecture but does not possess the authorized hardware configuration. The adversary's goal is to execute the model on unauthorized hardware with minimal performance degradation. Our locking mechanisms are not designed to protect against adversaries with physical access to the authorized hardware or those capable of sophisticated side-channel attacks. Defending against stronger adversaries would require extending our techniques to leverage existing security measures such as trusted execution environments.

\subsubsection{Example usage scenarios}

In medical AI, hardware locking could help prevent unauthorized access to models trained on sensitive patient data and ensure that diagnostic models are used only on approved, calibrated devices for accurate results. For instance, a diagnostic model could be locked to a specific MRI machine with a unique hardware fingerprint. In autonomous vehicles, hardware locking would allow new versions of models to be distributed freely with less risk to IP, and help ensure that safety critical models are not subtly tampered with by those without full access to the authorized vehicle hardware.

\section{Methodology}

\subsection{Assumptions, Goals, and Definitions}
\label{sec:method:assumption}

The goal of this paper is to develop \textit{ML Hardware Locking} mechanisms that make it hard to move a machine learning model from an authorized hardware platform to an unauthorized (different) hardware platform. That is, to build mechanisms that deter unauthorized model use by restricting supported hardware. Note that our locks are not designed to replace existing cryptographic security solutions e.g.\@ distribution and storage of encrypted weights, hardware security modules, and root of trust, but are rather developed to complement them for settings where restriction to specific hardware is preferred. Our methods are similar in function to standard encryption techniques, but use hardware-specific characteristics as key material, do not require explicit key management, and do not require specialised cryptographic hardware.

In what follows, we explicitly separate two main types of locking: hard and soft. \textbf{Soft locking} mechanisms refer to mechanisms that do not fully restrict normal use of the model on non-authorized hardware, but rather make use of the model less performant or efficient. For example, consider a model that during inference produces abnormally large amounts of internal data that slows down inference on normal GPUs, but specialised GPUs filter the produced data to only keep task-informative data. \textbf{Hard locking} mechanisms refer to mechanisms that fully restrict use of models on non-authorized hardware, ideally with formal e.g.\@ cryptographic guarantees. A mathematical formalism of these definitions can be seen in~\Cref{appendix:definitions}. In practice, we envision both locking types to be used in conjunction. To make such mechanisms usable, we seek to minimise effort required to instrument the model for deployment. At the same time, given that both developing proprietary hardware platforms and training large-scale foundation models are expensive, we either opt to \circled{\footnotesize{1}} convert models into representations that are unique to specific platforms e.g.\@ in~\Cref{tab:related_works} we show that \textbf{INT4} representation can be favourable for soft locking, since only two widely available hardware platforms support it; or \circled{\footnotesize{2}} condition models on hardware-specific behaviours e.g.\@ introducing model dependency on latency of scatter/gather operations.

We explicitly note that \textbf{none of the mechanisms described in this paper on their own provide security} and could be circumvented by a knowledgeable attacker with enough effort and appropriate access. Furthermore, they in no way prevent an adversary from performing model extraction, but these would result in significant adversary costs~\citep{tramer2016stealingapi,truong2021data,shafran2023labeling}. Yet, our locks present a challenge for an attacker with locked model-only access.

\subsection{Soft Locking Methods}
\label{sec:method:soft}
Optimisations for model inference, such as pruning and quantization, typically convert model parameters into sparse or low-precision tensors. These perturbations in the parameter space can lead to a test-time disparity, meaning that models, even if derived from the same original model but quantized with different arithmetics or pruned to varying sparsities, can misclassify distinct samples. Prior work has leveraged artefacts from quantization or pruning to develop stealthy backdoor attacks~\citep{ma2023quantization,hong2021quantization}. Nonetheless, in \textit{soft locking}, our interest lies in implementing strategies that enable optimisations \textit{exclusively} on hardware platforms with specific intrinsic support, not on others.
Unsupported or unauthorized hardware platforms may still be capable of running the same model but would suffer from inefficient execution and/or considerable performance degradation.

\subsubsection{Sparsity-aware locking} Pruning is a family of methods that transforms models with dense parameters into sparse ones \citep{lecun1989optimal}. Pruning reduces the number of parameters needed to store the model and potentially decreases the amount of floating-point operations required, if supported by the hardware. \Cref{tab:related_works} reveals that only NVIDIA A100, H100, Cerebras WSE2, and AMD MI300 possess hardware intrinsic support for sparse tensor acceleration, indicating the existence of hardware asymmetry in pruning. We propose a simple \textit{fine-tuning} scheme to produce models that significantly degrade in performance when used at an unauthorized sparsity level. The loss for this proposed manipulation takes the form: $\mathcal{L} = \mathcal{L}(f_p(x, p_1), y) + \lambda (\epsilon - \mathcal{L}(f_{p}(x, p_2), y))^2$. Here $f_{p}(\cdot)$ represents the pruned version of the original network $f(\cdot)$, where $x$ is the input and $p_1$ and $p_2$ are values between $0$ and $1$ that define the level of pruning in the pruned network. When $p=0$, we have an unpruned, dense network.
$\mathcal{L}$ denotes a valid loss, for example cross-entropy. $\lambda$ is a hyperparameter that allows tuning of the relative magnitudes of the original training term and the pruning-degradation term. $\epsilon$ denotes a target loss value for the pruned model.
In essence, our loss promotes free optimisation for models with a sparsity $p_1$ while aiming to drive models with a sparsity $p_2$ toward a suboptimal point, given that $p_1 \neq p_2$. This results in a sparsity-aware locking, where the model exhibits higher accuracy at a sparsity level of $p_1$ and significantly worse performance at a sparsity level of $p_2$.

\subsubsection{Quantisation-aware locking}
Machine Learning models are often distributed at lower quantization to allow deployment on specialist or constrained hardware. ML locking could therefore be accomplished by limiting which quantizations a particular model could be run at. We propose using a loss similar to the pruning-degradation loss defined above and that used by~\citet{hong2021quantization} to backdoor models, in order to lock the quantization levels which a model can be used on. The differences between our aim and \cite{hong2021quantization} are two-fold. Firstly, we aim to make transferring across hardware systems more challenging in order to accomplish ML locking instead of trying to backdoor model quantization. Secondly, we expand upon the approach of \cite{hong2021quantization} across various arithmetic types, not just within the integer arithmetic domain, as many chips listed in \Cref{tab:related_works} support integer arithmetic but vary in their support for different floating-point or even custom arithmetics.

\subsection{Hard Locking Methods}\label{sec:method-hard}

Our \emph{hard locking} methods are based on using a fingerprint obtained from a specific device to transform model parameters in a way that is difficult to invert without the fingerprint. Note that such signatures can be shared across devices from the same family, as we show with clock fingerprints, as well as, specific individual devices as we show with finite precision fingerprints.

First, the properties of the target device model are used to generate a \textit{high-entropy fingerprint} unique to the device or the device model, which is hard or impossible to replicate from other devices or device families. Then, a \textit{parameter transformation function} is used to modify the model parameters based on the fingerprint. The model is only ever stored in its transformed form, and detransformed on the fly by the authorized device. Without knowledge of the fingerprint, the transformed model is not functional, and the fingerprinting method is designed to have high entropy, such that it cannot easily be guessed or brute-forced without access to the authorized hardware.

\subsubsection{Device fingerprinting} The fingerprinting method can be anything which produces a consistent and unguessable output on one device or device model but a different output on other devices. It should have sufficient entropy to produce sufficient key material for the encryption-like parameter transformations. We propose three candidate device fingerprints: the \emph{clock fingerprint}, the \emph{finite precision fingerprint}, and the \emph{PUF fingerprint}. Different methods can be combined together for additional entropy.

\begin{table}[hbt]
\centering
\caption{Clock fingerprint on various devices.\\Further tests can be seen in~\Cref{sec:clock-fingerprint-data}}
\begin{tabular}{lll}
\toprule
	\textbf{Device}      & \textbf{GPU}    & \textbf{Fingerprint}     \\
	\midrule
	Tesla P100  & GP100  & 71900 to 7191f  \\
	GTX 1080Ti  & GP102  & 72100 to 723ff  \\
	RTX 2080Ti  & TU102  & 49a59           \\
	RTX 8000    & TU102  & 49a59           \\
	RTX 3090    & GA102  & 4b85a           \\
	RTX A6000   & GA102  & 4b85a           \\
	A100        &        & 4b059           \\
	GH200       &        & 4787c           \\
 \bottomrule
\end{tabular}

\label{tab:clock-fingerprint}
\end{table}

\subsubsection{Clock fingerprint} The \emph{clock fingerprint} is generated by counting the number of clock cycles taken by a CUDA device to repeatedly add numbers together. The fingerprint is this number represented as five symbols of hexadecimal. This can be seen in~\Cref{tab:clock-fingerprint}. We find experimentally that some devices such as the GP102 the fingerprint is stochastic while on others it is deterministic. The entropy of the clock fingerprint is upper bounded by the number of bits in the output (20). Clock fingerprints are an example of device-family fingerprint. We show our clock fingerprint generator in~\Cref{sec:clock_fingerprint_code}.

\subsubsection{Finite precision fingerprint} The \emph{finite precision fingerprint} is generated via numerical errors specific to an arithmetic and precision. In ML systems, these can also occur due to inference-time microbenchmarking \citep{schlogl2024causes}. Device-specific ML framework implementation choices produce a unique error which can be used as a fingerprint. The total entropy is determined by the total number of possible sets of such choices. \citet{schlogl2024causes} find a maximum of four error equivalence classes, or two bits of entropy, for a single convolution. The total bits of entropy available in a convolution-based finite precision fingerprint is therefore upper bounded by twice the number of convolutions performed. In reality, a large number of bits would be difficult to obtain just from convolutions due to strong correlations between the algorithm choices. At the same time, we find that floating point operations in general can be used to generate device-specific errors. We compute a finite precision fingerprint based on the SHA-256 hash of the error generated in a sequence of linear layers. These can be seen in~\Cref{sec:finite-precision-fingerprint-data}. We find that the fingerprint is consistent between different GPUs of the same model with the same framework version, but differ between different GPU models and framework versions. This could be used to further lock models to other components of the system. We show our finite precision fingerprint generator in~\Cref{sec:finite_precision_code}.

\subsubsection{PUF fingerprint}
Physically Unclonable Functions (PUFs) that exist in (or are explicitly introduced into) hardware can be used to derive a high-entropy device fingerprint. For example, \citet{vanaubel2015investigating} finds that the initial state of shared SRAM memory in NVIDIA GPUs is one example of an un-advertised PUF, while other constructions are possible~\citep{fengjun2015freqpuf,forlin2020g}. These PUFs could be combined with fuzzy extractors \citep{Dodis_2008} to achieve a reliable fingerprint. In model-distributed setups, these PUFs could be chained together across multiple devices. By choosing which bits of the SRAM are used for the PUF, the SRAM PUF fingerprint could be configured device-family level, device-model level, or even locked to a specific individual instance of the device.

\subsubsection{Parameter transformation} In order to achieve \emph{hard locking}, we must transform the model parameters in a way that can only be easily inverted with knowledge of the fingerprint. There are three ideal properties of the transformation, defined informally as follows:

\begin{enumerate}

    \item
    \textbf{Destruction:} The performance of a model with transformed parameters or parameters detransformed with an incorrect fingerprint should be equivalent to random guessing. Without destruction, the model would still be usable without the fingerprint, and hard locking fails.

    \item
    \textbf{Encryption:} No information about the original parameters should be obtainable from the transformed parameters, other than the information required for \emph{indistinguishability}. If encryption holds, then the attacker must brute-force all possible fingerprints to determine the correct one to reveal the secret. If it does not, in some cases cheaper methods such as gradient descent may be applicable for extracting the original parameters.

    \item
    \textbf{Indistinguishability:} Incorrectly detransformed parameters should be statistically indistinguishable in aggregate from correct parameters. If indistinguishability holds, then in a brute-force attack the attacker must run each candidate model on test data and choose the correct one by maximising test accuracy. This is in general more expensive and error-prone than a statistical test.

\end{enumerate}

\begin{table*}[hbt]
\caption{Comparison of parameter transformations that can be applied with different fingerprints.}

\centering
\begin{tabular}{lccc}
\toprule

  \textbf{Method}
& \textbf{Indistinguishablity}
& \textbf{Encryption}
& \textbf{Destruction} \\

\midrule

AES encryption                  & \xmark  & \cmark  & \cmark \\
Parameter shuffling             & \cmark  & \xmark  & \cmark \\
Pre-transformed AES encryption  & \cmark  & \cmark  & \cmark \\

\bottomrule
\end{tabular}
\label{tab:transforms}

\end{table*}

\Cref{tab:transforms} considers three parameter transformation methods in terms of these properties: \emph{AES encryption}, \emph{Parameter shuffling}, and \emph{Pre-transformed AES encryption}. These are described in detail below.

\subsubsection{AES encryption} The most obvious method to transform the model parameters is with a classical encryption scheme. This gives rise to the \emph{AES encryption} method, in which the parameters of the model are collected together into a bytestream, which is then AES-encrypted with the SHA-256 hash of the fingerprint used as the key. The resulting bytestream is then interpreted as transformed parameters. This achieves perfect \emph{encryption} and \emph{destruction}, but not \emph{indistinguishability}, because incorrectly decrypted parameters will be uniformly distributed, in contrast to the correct parameters.

In conventional cryptographic schemes, security can be increased by using higher-entropy keys to make brute-forcing infeasible. Since the entropy of a fingerprint is fixed and cannot be increased, we focus instead on making each decryption attempt more expensive, i.e. key-stretching \citep{kelsey1997secure}. We propose achieving this through \emph{indistinguishability}. With indistinguishability, for each candidate fingerprint the attacker must evaluate the candidate model's accuracy, which is computationally expensive. Indistinguishability of parameters is feasible since the statistics of ML parameters are much easier to fake than the plaintext of most encryption schemes (for example, coherent English text, or a valid JPEG file).

In order to achieve \emph{indistinguishability} and thus use key stretching to make a brute-force attack more difficult, we developed two further transformation methods: \emph{parameter shuffling}, and \emph{pre-transformed AES encryption}.

\subsubsection{Parameter shuffling} In the \emph{parameter shuffling} method, the fingerprint is used as a seed to generate a random permutation of the parameters, which is then used in place of the original parameters. This achieves near-perfect \emph{indistinguishability} and \emph{destruction}, but does not achieve full \emph{encryption}, as some information about the original parameters still exists in the permuted parameters.

In practice however, because the search space of permutations ($n!$ where $n$ is the number of parameters) is many orders of magnitude larger than the search space of keys ($2^b$ where $b$ is the fingerprint entropy), it is much faster for an attacker to brute force the key space than attempt gradient descent in the permutation space, and so the lack of perfect encryption is unimportant. For example, a very large fingerprint with 256 bits of key material has a search space of $2^{256} \approx 10^{80}$, but even a small ResNet18 test model has over $10^7$ parameters, equating to a permutation search space of $\left(10^7\right)! \approx 10^{{10}^8}$.

\subsubsection{Pre-transformed AES encryption} The problem with na\"{\i}ve AES transformation is that incorrectly detransformed parameters are uniformly distributed in byte-space, while correct parameters have some other distribution, often Gaussian, making cracking considerably easier.

We can correct for this problem by defining an additional transformation function, the `pre-transformation' function. This is applied to the parameters before encryption, resulting in uniform bytes very difficult to statistically distinguish from incorrectly decrypted bytes. The reversed pre-transformation function is stored unencrypted with the model, and applied to these uniform bytes after decryption is attempted, resulting in what appear to be valid parameters regardless of whether the key was correct. We therefore say that \emph{pre-transformed AES encryption} achieves the full trifecta of \emph{indistinguishability}, \emph{encryption}, and \emph{destruction}.

Any distribution can be made uniform by applying its cumulative distribution function to it. For example, if $X$, the distribution of model parameters, is Gaussian with mean $\mu$ and variance $\sigma^2$, then $Y$, the distribution of pre-transformed model parameters, is uniform in the region $(0, 1)$, where:

\begin{equation*}
Y = \frac{1}{2} \left[1 + \textrm{erf}\left(\frac{X - \mu}{\sigma\sqrt{2}}\right)\right].
\end{equation*}

If $Y$ is then encoded as integers spanning the full possible integer range, its distribution will also be uniform in byte space, and thus indistinguishable from incorrectly decrypted bytes.

We evaluated pre-transforming the parameters of a test ResNet18 model trained on CIFAR10, first by assuming that it was Gaussian, and secondly by directly estimating the cumulative distribution function via sampling of the parameters. This can be seen in \Cref{sec:transform-distributions}.

In practice, assuming that the distribution is Gaussian may be problematic. If there are any outliers, then applying the cumulative distribution function will take these outliers so close to 0 or 1 that they can no longer be represented with sufficient precision in floating point, and they will become exactly 0 or 1. Then, these outliers detransform to positive or negative infinity, destroying the model. There are sufficiently many outliers that subsequently casting these outliers to any fixed finite value will destroy the accuracy of the model. Regardless of any scheme to correct for this, \emph{indistinguishability} is broken, because these infinities do not occur with comparable frequency in incorrectly detransformed data. To fix this problem, we computed the empirical distribution of the parameters and used a look-up table to transform this into a uniform distribution. For an $n$-bit precision, computing the pre-transformation requires $\mathcal{O}(2^n)$ time and space, but subsequently applying the pre-transformation or reversed pre-transformation is cheap. This is computationally tractable for precisions up to 32-bits and inexpensive for precisions up to 16-bit. The results for FP16 can be seen in~\Cref{fig:transform-distributions-pretransformed-direct-encrypt} in \autoref{sec:transform-distributions}. There exists a trade-off between the pre-transform being exactly invertible and it exactly converting uniform data into the target distribution data. Reported results are for an exactly invertible pre-transformation.

\section{Evaluation}
\label{sec:evaluate}
\subsection{Evaluating Soft Locking}

We evaluate the effects of different soft locking schemes as illustrated in \Cref{tab:taxonomy}.
When run on unauthorized hardware, soft-locked models will cause either a slow down in inference, or a drop in accuracy or model performance. The former occurs because unauthorized hardware may emulate the execution of authorized hardware at the software level, due to a lack of hardware intrinsic support, this will result in slowdowns, as we evaluate in \Cref{sec:eval:soft:cost}. Alternatively, if unauthorized hardware directly executes only what is natively supported, it would incur an accuracy penalty, as illustrated in \Cref{sec:eval:soft:sparsity}, and \Cref{sec:eval:soft:quantization}.

\subsubsection{Performance degradation for sparsity-aware lock}
\label{sec:eval:soft:sparsity}

We investigate sparsity-aware locking, for the widely-used $l_{1}$-unstructured pruning \citep{lecun1989optimal, li2016pruning} across pruning levels (0.05, 0.10, 0.25, 0.50, 0.75) for both vision and language models, across different datasets. For all vision models, we fine-tune a trained model \footnote{Training and fine-tuning hyper-parameters presented in the Appendix \ref{hparams}.} with the loss defined in \Cref{sec:method:soft} for 25 epochs, with $\lambda$ = 1 and $\epsilon$ = 5. The language models are fine-tuned for 3 epochs, as this was sufficient for effective manipulation.

We fine-tuned the BERT model \citep{devlin2018bert}, and specifically the \textit{bert-base-cased-finetuned-\{sst2, cola, mrpc\}} HuggingFace checkpoints on the respective GLUE tasks \citep{wang2018glue}, with the pruning-resistant loss from~\Cref{sec:method:soft}. We also evaluate performance on various vision tasks such as CIFAR10, CIFAR100 \citep{krizhevsky2014cifar}, and Flowers102 \citep{nilsback2008automated}, utilising ResNet18, ResNet50 \citep{he2016deep}, and ViT-B \citep{dosovitskiy2020image} for these specific tasks. For Flowers102, we trained on the 6149 image `test' dataset rather than the 1020 image `train' dataset, and evaluated on the 1020 image `val' dataset, inline with modern training/evaluation dataset split sizes.

\begin{figure}[t]
	\includegraphics[width=\linewidth]{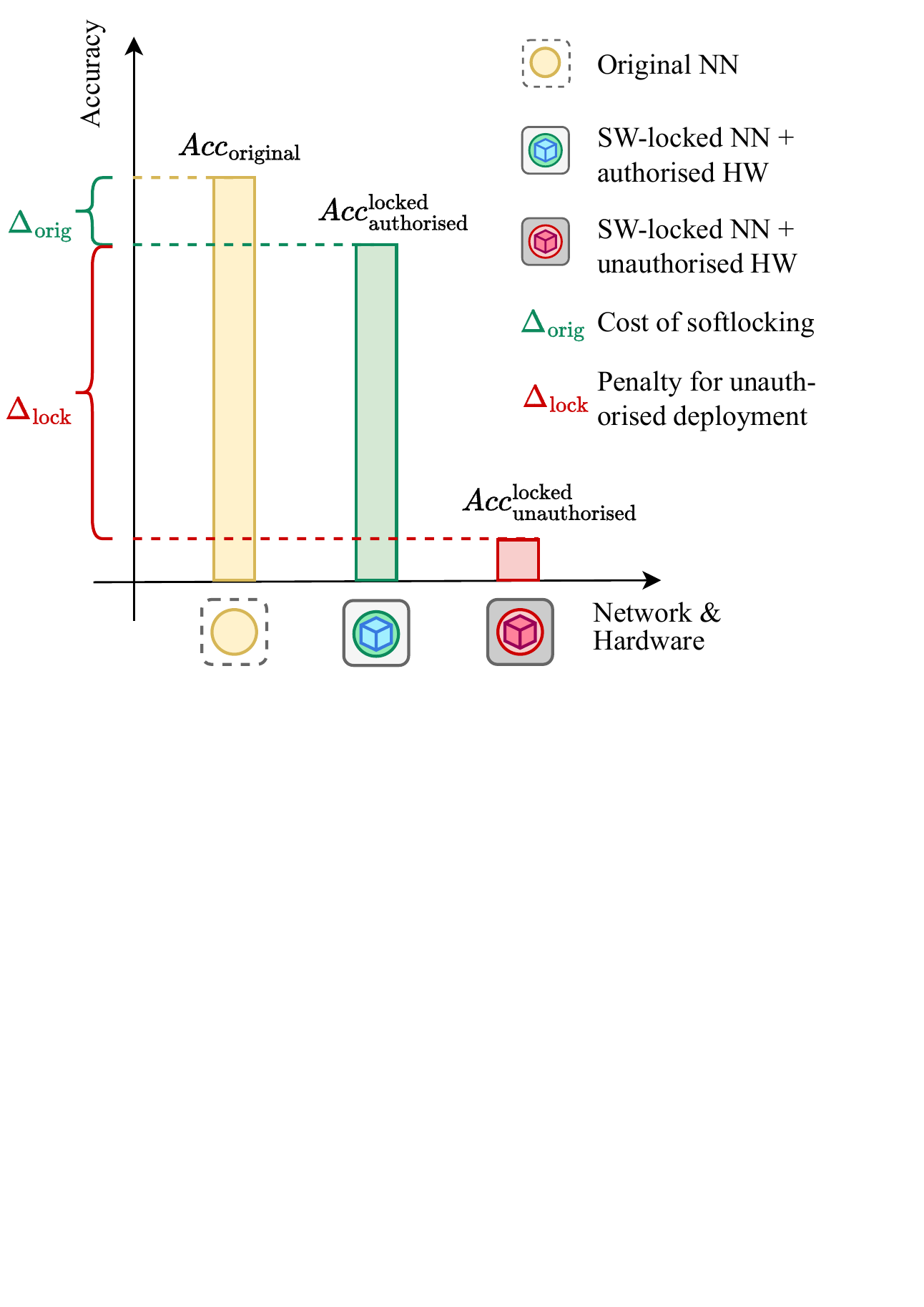}
	\vspace{1.1mm}
    \caption{Applying soft locks to a model with $\mathit{Acc}_\text{original}$, $\textcolor{czgreen}{\Delta_{\text{orig}}}$ and $\textcolor{czred}{\Delta_{\text{lock}}}$ measure together the effectiveness of locking.}
    \label{fig:terminology}
\end{figure}

\Cref{tab:pruning} displays sparsity-aware lock results, we present the following metrics as illustrated in \Cref{fig:terminology}:

\begin{itemize}
    \item $\mathit{Acc}_{\text{authorized}}^{\text{locked}}$: the accuracy of models with soft locks running on authorized hardware.
    \item $\textcolor{czgreen}{\Delta_\text{orig}} = \mathit{Acc}_{\text{original}} - \mathit{Acc}_{\text{authorized}}^{\text{locked}}$ measures the impact of soft locking on authorized execution by comparing the accuracy of the original model (without locks) to that of the locked model executing on authorized hardware. A \textbf{small} $\textcolor{czgreen}{\Delta_\text{orig}}$ value is desirable.
    \item $\textcolor{czred}{\Delta_{\text{lock}}} = \mathit{Acc}_{\text{authorized}}^{\text{locked}} - \mathit{Acc}_{\text{unauthorized}}^{\text{locked}}$ measures the degradation in accuacy when locked models are deployed on unauthorized devices. A \textbf{large} $\textcolor{czred}{\Delta_\text{lock}}$ value is desirable.
\end{itemize}

\Cref{tab:pruning} demonstrates that sparsity-aware lock leaves the accuracy of the authorized execution largely unaffected, as shown by the small $\textcolor{czgreen}{\Delta_{\text{orig}}}$ values. Meanwhile, locked models experience significant accuracy degradation when operating on unauthorized hardware, as indicated by the substantial $\textcolor{czred}{\Delta_{\text{lock}}}$ values. We present the accuracy drop from hardware transfer without locking, $\Delta_{\text{base}}$, in the Appendix (\ref{soft-base}). It is clear that the significant accuracy drop from transfer can be attributed to locking (and not merely pruning), as the degradation in accuracy of the original models, $\Delta_{\text{base}}$, is generally much lower than that of the locked models, $\textcolor{czred}{\Delta_{\text{lock}}}$. Certain settings, such as \textit{BERT} on MRPC, show that the sparsity-aware lock can trigger a larger degradation in performance when the unauthorized configuration is more sparse (i.e. greater pruning proportion, $p$). However, even a low unauthorized pruning proportion of $p = 0.05$ is sufficient for sparsity-aware locks to be effective.

\begin{table}[t]
\caption{Results are presented as $\mathit{Acc}_{\text{authorized}}^{\text{locked}} (\textcolor{czgreen}{\Delta_\text{orig}}, \textcolor{czred}{\Delta_{\text{lock}}})$.
}
\centering
\adjustbox{max width=\linewidth}{
\begin{tabular}{lccc}
\toprule
 & \multicolumn{3}{c}{\textbf{Pruning Levels}} \\
 \cmidrule(lr){2-4}
 \textbf{Dataset} & 0.05 & 0.25 & 0.50 \\
\midrule
 & \multicolumn{3}{c}{\textit{BERT}} \\
CoLA & 0.80 (-0.03, 0.49) & 0.84 (0.00, 0.53) & 0.83 (0.00, 0.53) \ \\
MRPC & 0.84 (-0.02, 0.15) & 0.86 (0.00, 0.18) & 0.86 (0.00, 0.54) \ \\
SST-2 & 0.92 (-0.01, 0.43) & 0.93 (0.00, 0.43) & 0.92 (0.00, 0.43)\ \\
\midrule
 & \multicolumn{3}{c}{\textit{Resnet18}} \\
CIFAR10 & 0.89 (0.03, 0.67) & 0.93 (0.00, 0.83) & 0.93 (0.00, 0.83) \ \\
CIFAR100 & 0.73 (0.03, 0.71) & 0.76 (0.00, 0.75) & 0.76 (0.00, 0.75)\ \\
Flowers102 & 0.84 (0.04, 0.83) & 0.89 (0.00, 0.88) & 0.89 (0.00, 0.88) \ \\
\midrule
 & \multicolumn{3}{c}{\textit{Resnet50}} \\
CIFAR10 & 0.92 (0.01, 0.81) & 0.93 (-0.01, 0.83) & 0.94 (-0.01, 0.84) \ \\
CIFAR100 & 0.69 (0.09, 0.63) & 0.78 (0.00, 0.77) & 0.78 (0.00, 0.77) \ \\
Flowers102 & 0.76 (0.10, 0.74) & 0.86 (0.00, 0.84) & 0.86 (0.00, 0.84) \ \\
\midrule
 & \multicolumn{3}{c}{\textit{ViT-B\_16-224}} \\
CIFAR10 & 0.99 (0.00, 0.89) & 0.99 (0.00, 0.89) & 0.99 (0.07, 0.89) \ \\
CIFAR100 & 0.92 (-0.02, 0.92) & 0.93 (-0.02, 0.92) & 0.93 (-0.02, 0.93) \ \\
Flowers102 & 1.00 (0.00, 0.98) & 1.00 (0.00, 0.99) & 1.00 (0.00, 0.99) \ \\
\bottomrule
\end{tabular}}
\label{tab:pruning}
\end{table}

\subsubsection{Performance degradation for quantization-aware lock}

\label{sec:eval:soft:quantization}
We investigate quantization-aware locking on ResNet models across a set of authorized, unauthorized arithmetic pairs, using the quantization-aware locks described in \Cref{sec:method:soft}. The models are trained for 25 epochs, with $\lambda$ = 1 and $\epsilon$ = 5, as with sparsity-aware locking. We note that \cite{hong2021quantization} previously used a similar loss in~\Cref{sec:method:soft} to learn models that degrade upon quantization to a lower precision. We extend this by investigating the effectiveness of preventing transfer across both precision and arithmetics and present the results below, using the metrics from \Cref{sec:eval:soft:sparsity}.

\begin{table}[b]
\caption{Results are presented as $\mathit{Acc}_{\text{authorized}}^{\text{locked}} (\textcolor{czgreen}{\Delta_\text{orig}}, \textcolor{czred}{\Delta_{\text{lock}}})$. %
}
\centering
\begin{tabular}{llll}
\toprule
 Authorized $\rightarrow$ Unauthorized & \textbf{Resnet18} & \textbf{Resnet50} \\
\midrule
FP32 $\rightarrow$ 8-bit MiniFloat
& 0.90 (-0.02, 0.65) & 0.92 (-0.04, 0.67) \\
Int8 $\rightarrow$ 8-bit MiniFloat & 0.47 (+0.41, 0.06) & 0.50 (+0.39, 0.07) \\
FP16 $\rightarrow$ Int8 & 0.90 (-0.02, 0.62) & 0.91 (-0.04, 0.61) \\
16-bit MiniFloat\textsuperscript{\ref{footnote:minifloat}} $\rightarrow$ Int8 & 0.90 (-0.01, 0.72) & 0.91 (-0.03, 0.81) \\
\bottomrule
\end{tabular}
\label{tab:quan}
\end{table}

Here, we use FP32 to simulate these formats during fine-tuning to prove the idea, which is independent of hardware. In both models, the quantization-aware lock causes performance to degrade to near-random guessing performance when the quantization is \textit{to a lower precisions than the original model}, as evident by both \textbf{FP32} to \textbf{8-bit MiniFloat }and \textbf{16-bit MiniFloat}\footnote{\label{footnote:minifloat}Besides FP16, vendors have various MiniFloat formats, such as Google's \textbf{BFloat16} and NVidia's \textbf{TensorFloat}. We set exponent width 5 and bias 11.} to \textbf{Int8} in \Cref{tab:quan}. Manipulation across the same precision, but different arithmetic, was not successful (\textbf{Int8} to \textbf{8-bit MiniFloat}), showing there is not sufficient discrepancy in their representations for this simple manipulation procedure to yield model degradation. However, the transfer across arithmetics (\textbf{16-Bit MiniFloat} to \textbf{Int8}) results in a greater degradation, with a lower difference in precision than transfer across just precision (\textbf{FP32} to \textbf{8-bit MiniFloat}). %

\subsubsection{Attacking soft locking via re-training}
\label{appendix:attacking_vision}

If the unauthorized user has access to data, they may \textit{attack} the soft locking procedures by re-training the models on the unauthorized hardware. For re-training, it is natural for the user to employ a quantization- or pruning-aware loss and minimise the loss of the quantized or the pruned model.

\begin{figure}[!b]
    \centering

    \includegraphics[trim={0 0 0 12mm},clip,width=\linewidth]{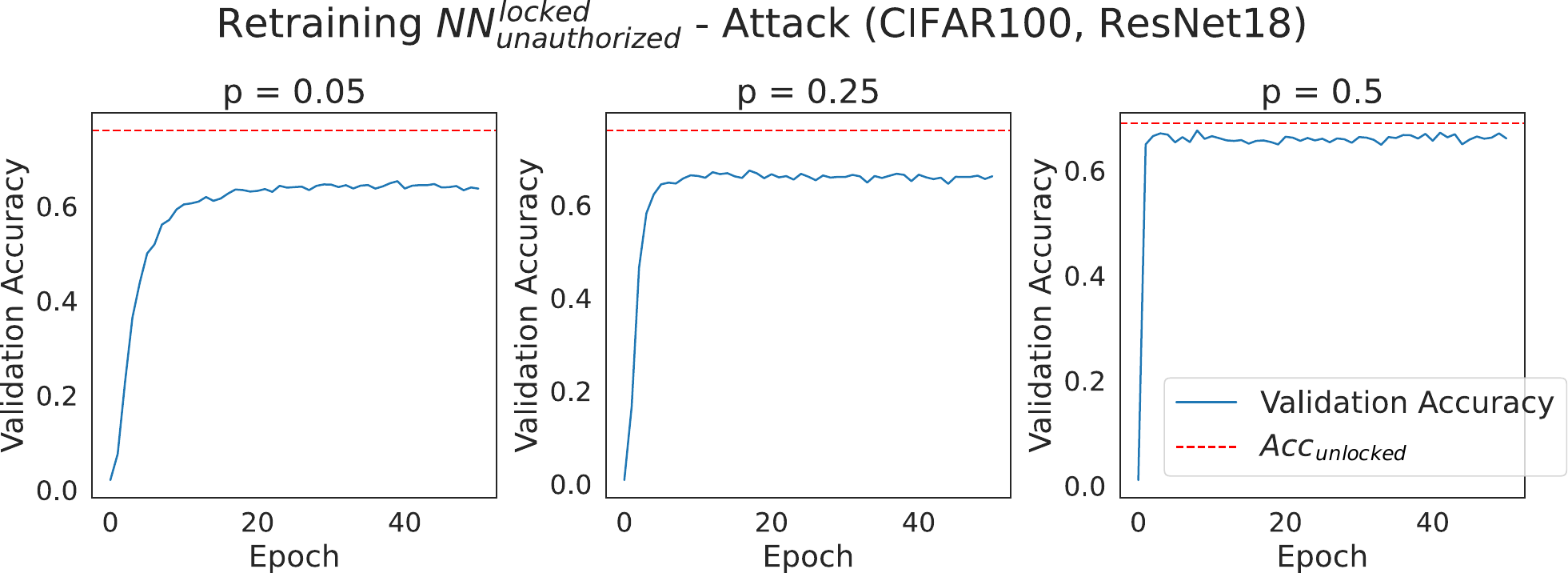}

    \vspace{-0.2\baselineskip}
    \caption{Re-training sparsity-locked \textit{ResNet18} on CIFAR100}
    \label{fig:resnet18_retraining}
    \vspace{\baselineskip}

    \includegraphics[trim={0 0 0 12mm},clip,width=\linewidth]{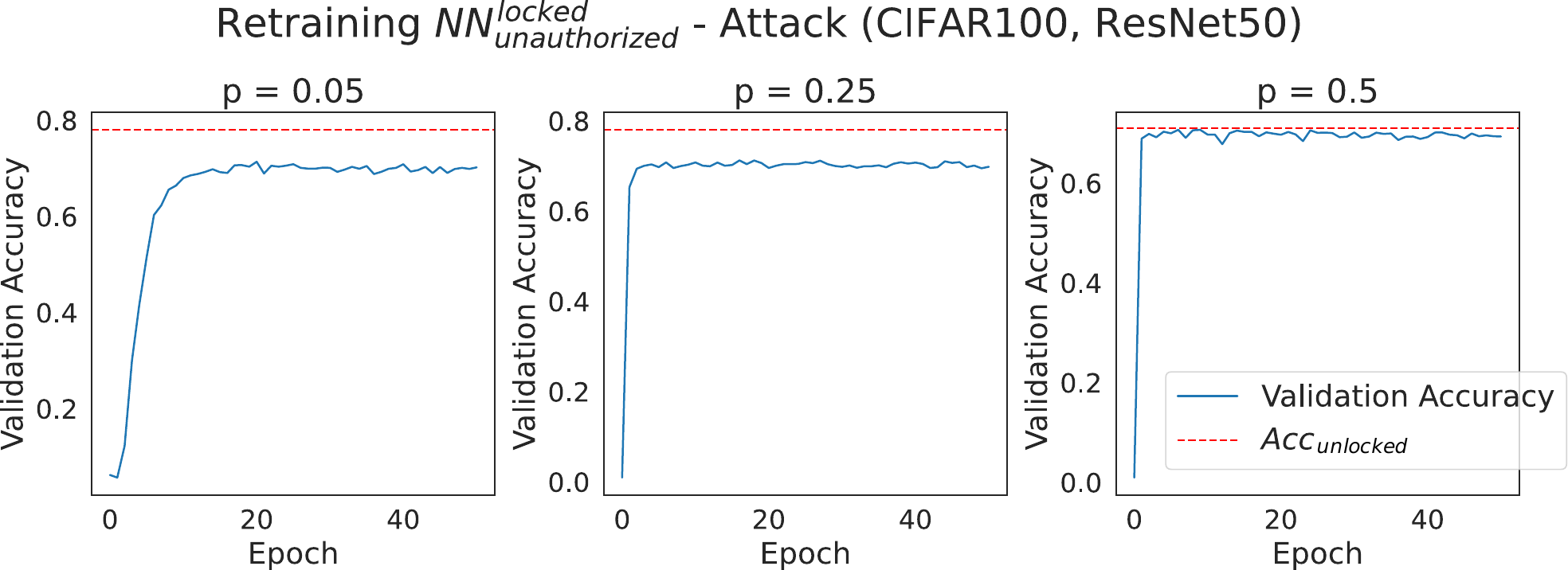}

    \vspace{-0.2\baselineskip}
    \caption{Re-training sparsity-locked \textit{ResNet50} on CIFAR100}
    \label{fig:resnet50_retraining}
    \vspace{\baselineskip}

     \includegraphics[width=\linewidth]{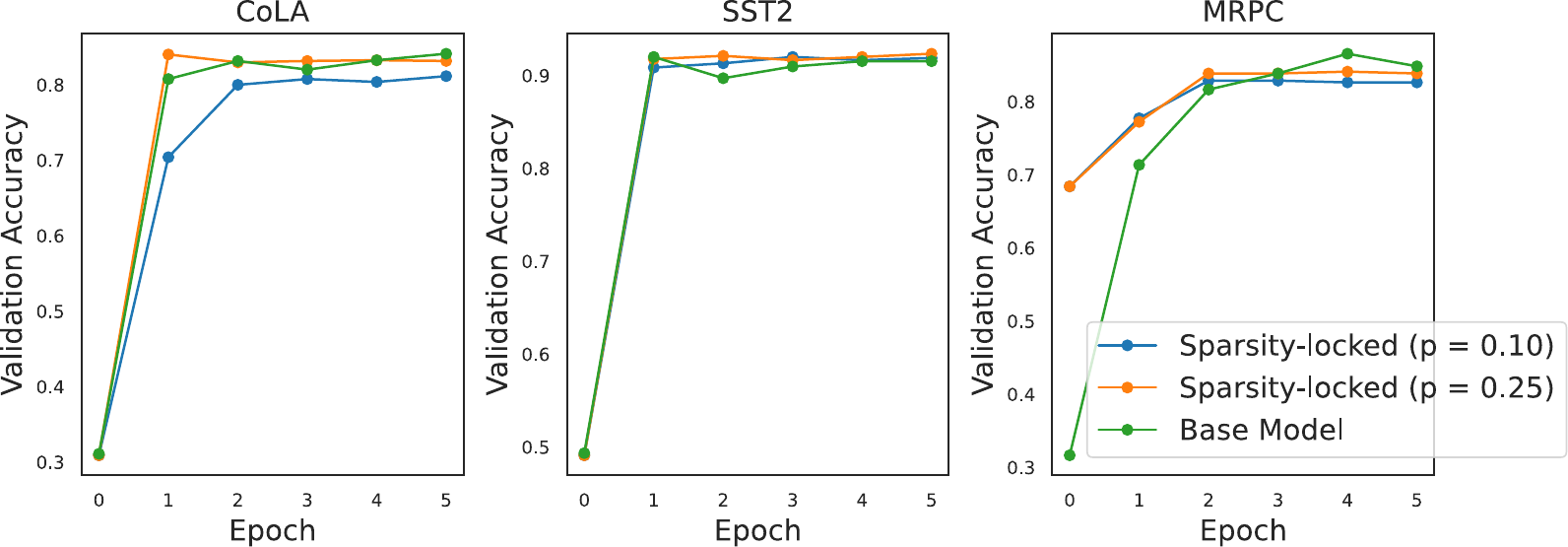}

     \vspace{-0.2\baselineskip}
     \caption{Re-training sparsity-locked \textit{BERT} on GLUE tasks}
     \label{fig:retraining_llm}
\end{figure}

To investigate the effectiveness of re-training, we re-trained a subset of the soft-locked vision models. This can be seen in~\Cref{fig:resnet18_retraining,fig:resnet50_retraining}. We find that at lower levels of sparsity, $p$ (0.05 and 0.25), re-training does not allow the model to reach the accuracy of the original (i.e. unlocked) model running on unauthorized hardware. At $p = 0.50$, accuracy is regained after roughly 5 epochs. As such, the locking procedure is highly effective, as in the best case it causes damage that cannot easily be recovered from, and either way it necessitates further training from the unauthorized users, who may not have access to the training data. We present below the re-training curves of the sparsity-locked models.

In both models, re-training the models locked at $p = 0.05$ and $p = 0.25$ does not return them to the accuracy of the unlocked model deployed on unauthorized hardware. For the models locked at $p = 0.50$, the majority of the accuracy is recovered in roughly 5 epochs.

In~\Cref{fig:retraining_llm} we also present re-training the sparsity-locked \textit{BERT} models on the GLUE tasks, with a pruning-aware loss. The fine-tuning curve of a base \textit{BERT} model is presented as a baseline of reference. The recovery of performance of locked models is comparable to the fine-tuning a sparse \textit{BERT} model from scratch. However, we note that GLUE tasks may not provide the resolution that the previous vision tasks did, as 5 epochs are sufficient for convergence in this setting.

We present further investigation in~\Cref{appendix:optimisation_profiles,appendix:recovery} into typical soft locking profiles and the specificity of soft locking to the assumed unauthorized sparsity level.

\subsubsection{Attacking soft locking via noise}\label{sec:noise_attack}

\begin{figure}[!b]
	\centering
	\includegraphics[width=\linewidth]{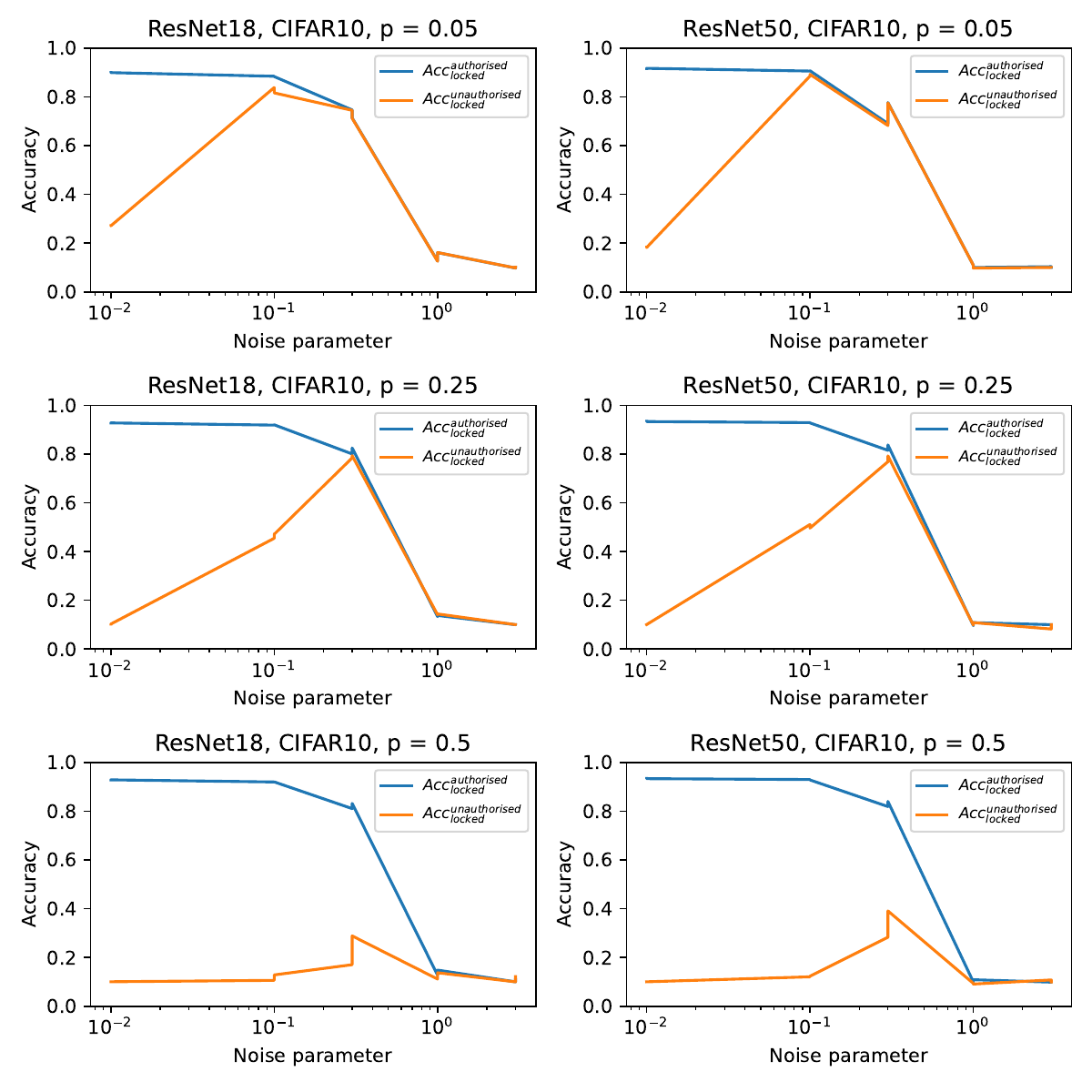}
	\vspace{-\baselineskip}
	\caption{Attacking soft locking by adding noise to the locked parameters.}
	\label{fig:noise_attack}
\end{figure}

We also investigated removing soft locking by adding noise to the parameters, based on the idea that the locked state may only be a very small region of parameter space. Results for the sparsity-aware lock for ResNet on CIFAR10 can be seen in~\Cref{fig:noise_attack}. Here, noise is added to each parameter vector in the model proportional to the standard deviation of the vector. The proportional factor is the \emph{noise parameter}. It is possible to see that in some setups, simply adding noise can recover significant performance on unauthorized hardware, and in other setups, hardly any performance can be recovered. Adding noise is much cheaper than re-training, so hyper-parameters must be carefully chosen in soft locking setups to avoid this attack.

\addtocounter{table}{1}
\begin{table*}[b]
\caption{
Accuracy of test model (FP16 ResNet18 trained on CIFAR10) with parameter
transformations. 10\% is random guessing. "Correctly detransformed" is
detransformed with the same fingerprint that was transformed with, while
"Incorrectly detransformed" is detransformed with any other fingerprint. Cost
refers to experimental cost of brute-forcing all possible fingerprints on the
test model, in CPU time. $b$ is the number of bits of entropy in the
fingerprint, discussed further in~\Cref{sec:hard-locking-entropy}.
}
\centering
\adjustbox{max width=\linewidth}{
\begin{tabular}{lccccc}
\toprule

~                & \multicolumn{4}{c}{\textbf{Accuracy}} & \textbf{Cracking cost (s)} \\
\cmidrule(lr){2-5}
\textbf{Method}  & \textbf{Original}  & \textbf{Transformed}  & \textbf{\makecell{Correctly\\detransformed}}  & \textbf{\makecell{Incorrectly\\detransformed}} & \\

\midrule

AES encryption                  & 95.4\%  & 10\%  & 95.4\%  & 10\% & $0.3 \times 2^b$  \\
Parameter shuffling             & 95.4\%  & 10\%  & 95.4\%  & 10\% & $2.7 \times 2^b$ \\
Pre-transformed AES encryption  & 95.4\%  & 10\%  & 95.4\%  & 10\% & $1.3 \times 2^b$  \\
\bottomrule
\end{tabular}}
\label{tab:transform-accuracy}
\end{table*}

\addtocounter{table}{-2}
\begin{table}[!t]
\caption{Emulation costs for soft locking. Though the accuracy of soft locked model can be recovered by emulation on unauthorized hardware, the inference suffers from inefficient execution, such as a lower throughput and higher latency. The single \texttt{matmul} is at a size of (2048, 2048), and we use the OPT-2.7B model at a batch size of 4 on NVIDIA A6000. TOPS denotes Tera operations per second, and TPS is tokens per second.}
\label{tab:emulation-cost}
\centering
\adjustbox{width=\linewidth}{
\begin{tabular}{llrrrr}
\toprule
\multicolumn{1}{c}{}
& \multicolumn{1}{c}{}
& \multicolumn{2}{c}{\textbf{Sparsity-aware}}
& \multicolumn{2}{c}{\textbf{Quantisation-aware}} \\
\cmidrule(lr){3-4} \cmidrule(lr){5-6}
\multicolumn{1}{l}{\textbf{Workload}}
& \multicolumn{1}{l}{\textbf{Metric}}
& \multicolumn{1}{r}{\textbf{Real}}
& \multicolumn{1}{r}{\textbf{Emulated}}
& \textbf{Real}
& \multicolumn{1}{r}{\textbf{Emulated}} \\
\midrule
\multirow{2}{*}{Single Matmul}
&  Throughput (TOPS)
&   49.97
&   18.85
&   79.22
&   22.72
\\
&  Latency (ms)
&   0.43
&   1.14
&   0.27
&   0.95
\\
\midrule
\multirow{2}{*}{OPT inference}
&  Throughput (TPS)
&  4692.20
&  2468.31
&  3505.22
&  1865.41
\\
&  Latency (ms)
&  436.47
&  829.72
&  584.27
&  1097.88
\\
\bottomrule
\end{tabular}
}
\end{table}

\addtocounter{table}{1}

\subsubsection{Emulation cost for soft locking}
\label{sec:eval:soft:cost}

As described in \Cref{sec:method:soft} and shown in \Cref{tab:taxonomy}, unauthorized devices without supported hardware intrinsics may opt for operation emulation, which incurs additional throughput costs. For a fair comparison, we assess the performance of both a single matrix multiplication operation (\texttt{matmul}) and the prefill phase of the entire network with and without software emulation on the same device, an NVIDIA A6000 GPU.
In \Cref{tab:emulation-cost}, for sparsity-aware locking, the term ``\textbf{Real}'' signifies that weight matrices exhibit a sparsity level of $0.995$, implemented via \texttt{torch.sparse.mm}. Conversely, ``\textbf{Emulated}'' indicates that the linear layer employs a full weight matrix alongside a $0.995$-sparsity mask; during runtime, this mask is applied to the weights, which are then processed using \texttt{nn.functional.linear}.
Regarding quantization-aware lock, ``\textbf{Real}'' means the hardware executes the quantized operation using its intrinsically supported INT8 GEMM, where ``\textbf{Emulated}'' is emulating the INT8 operations using ordinary FP32 operations.
We use the MASE flow for both sparsity and quantization emulations \citep{cheng2023fast}.
We consider both a single matrix multiply as the size of (2048, 2048) and a full OPT-2.7B model inference at a batch size of 4. %

Our results in \Cref{sec:method:soft} suggest that emulation comes at a large cost. For instance for a single matrix multiply, both sparsity-aware and quantization-aware locks induce a 2.65$\times$ and 3.52$\times$ overhead in latency, and a 2.65$\times$ and 3.49$\times$ reduction in throughput. For the inference of OPT-2.7B, the emulation overhead is around 2$\times$ in both latency and throughput.

\subsection{Evaluating Hard Locking}

We tested the three transformation functions presented in
\autoref{sec:method-hard} on ResNet18 \citep{he2016deep} trained on CIFAR10 \citep{krizhevsky2014cifar}. All three successfully achieve \emph{destruction}, as can be seen in~\Cref{tab:transform-accuracy}. We further tested the cost to fully brute force each. In our tests on this very small model, the \emph{indistinguishability} property of the \emph{Parameter shuffling} and \emph{Pre-transformed AES} methods added to the cracking cost significantly. The difference between the cost of \emph{Parameter shuffling} and \emph{Pre-transformed AES} can be attributed to AES being a more highly optimised transformation than shuffling. We expect the difference in cost between these two and the non-indistinguishable \emph{AES encryption} method will be much more significant with larger models, as will be discussed further in~\Cref{sec:scalability}.

\subsection{Cost and Scalability of Soft and Hard Locking}\label{sec:scalability}

Neither soft nor hard locking adds additional storage cost: in soft locking, once the locked model is generated the unlocked model is not needed, and in hard locking, the model is stored transformed and detransformed in memory as it is loaded.

Soft locking adds no compute cost to using the model regardless of model size. The compute cost of creating a soft-locked model scales with the cost of fine-tuning the model.

The compute cost of creating a hard locked model is the same as the cost of transforming the model parameters once, which is between 0.3 and 2.7 seconds of CPU time using unoptimized code for a small model as presented in~\Cref{tab:transform-accuracy} (equivalent to brute forcing with entropy $b = 0$). This should scale linearly with the number of parameters. This cost is incurred when using the model, but only when the model is loaded, not during inference.

In larger models than our small test model, the benefit of the
\emph{indistinguishability} property of the \emph{Parameter shuffling} and \emph{Pre-transformed AES} methods will increase significantly. This is because for a number of parameters $n$, while the cost of transformation is $O(n)$, the cost of inference is usually significantly superlinear with $n$, for instance anywhere matrix multiplication is involved. In other words, the brute force cost for the attacker will be dominated by the cost of running the model to test if decryption succeeded, rather than the cost of transformation. This enables strong key-stretching without impacting the cost of using the model: the transformation can be made more efficient without changing the brute force cost.

\section{Scope and Limitations}

Soft and hard locking mechanisms each present unique trade-offs in terms of security, flexibility, and performance. Soft locking, primarily based on quantization and sparsity, offers a lightweight approach suitable for scenarios where some level of model degradation on unauthorized hardware is tolerable. It is particularly useful when the primary concern is deterring casual misuse rather than sophisticated attacks. However, its effectiveness diminishes against determined adversaries who might employ techniques like fine-tuning to circumvent the lock. It is worth noting that access to the training data would be required, and the cost of fine-tuning to remove the lock can be similar to the cost of fine-tuning the model for the task in the first place. Some results of attacking via fine-tuning can be seen in~\Cref{appendix:attacking_vision}.

Hard locking, leveraging hardware-specific fingerprints, provides stronger security guarantees by binding the model to a specific hardware configuration. This makes it significantly more challenging for adversaries to execute the model on unauthorized hardware. Depending on the choice of fingerprint, a hard-locked model can locked to a device family, specific device model, or even a specific device. This gives a level of granularity to choose how widely accessible the model should be, based on hardware. However, hard locking can introduce complexities in deployment, especially when models need to be executed on a variety of hardware platforms.

Soft and hard locking are designed to complement existing security
methods, not replace them. They do not require explicit key management or dedicated cryptographic hardware, so offer a lightweight and potentially more accessible solution for ML model protection, especially where traditional trusted execution environments might be impractical or unavailable.

There are potential challenges with at-scale deployment, particularly with respect to minor chip revisions, device degradation, and driver updates. To address this, locks could allow for configurable tolerance levels to accommodate minor hardware variations without triggering a lockout, or rely on additional mechanisms such as hardware security modules.

\section{Conclusion}

In this paper we introduce \textit{ML Hardware Locking}, a novel paradigm for protecting machine learning models from unauthorized use, to address a growing concern about intellectual property protection and responsible AI use. We investigate a number of different locking mechanisms, encompassing both soft locking, which discourages model theft by imposing performance penalties on unauthorized hardware, and hard locking, which leverages hardware fingerprints to (cryptographically) bind models to specific platforms. Our experiments demonstrated the effectiveness of these locking mechanisms, both preserving model performance and introducing significant complexity in removing the locks. By investigating hardware-based locking mechanisms, we offer a potential solution for safeguarding valuable on-device machine learning models.

\FloatBarrier
\bibliographystyle{IEEEtranN}
\bibliography{references}

\begin{thebibliography}{48}
\providecommand{\natexlab}[1]{#1}
\providecommand{\url}[1]{#1}
\csname url@samestyle\endcsname
\providecommand{\newblock}{\relax}
\providecommand{\bibinfo}[2]{#2}
\providecommand{\BIBentrySTDinterwordspacing}{\spaceskip=0pt\relax}
\providecommand{\BIBentryALTinterwordstretchfactor}{4}
\providecommand{\BIBentryALTinterwordspacing}{\spaceskip=\fontdimen2\font plus
\BIBentryALTinterwordstretchfactor\fontdimen3\font minus
  \fontdimen4\font\relax}
\providecommand{\BIBforeignlanguage}[2]{{%
\expandafter\ifx\csname l@#1\endcsname\relax
\typeout{** WARNING: IEEEtranN.bst: No hyphenation pattern has been}%
\typeout{** loaded for the language `#1'. Using the pattern for}%
\typeout{** the default language instead.}%
\else
\language=\csname l@#1\endcsname
\fi
#2}}
\providecommand{\BIBdecl}{\relax}
\BIBdecl

\bibitem[Brown et~al.(2020)Brown, Mann, Ryder, Subbiah, Kaplan, Dhariwal,
  Neelakantan, Shyam, Sastry, Askell, Agarwal, Herbert-Voss, Krueger, Henighan,
  Child, Ramesh, Ziegler, Wu, Winter, Hesse, Chen, Sigler, Litwin, Gray, Chess,
  Clark, Berner, McCandlish, Radford, Sutskever, and Amodei]{brown2020language}
\BIBentryALTinterwordspacing
T.~Brown, B.~Mann, N.~Ryder, M.~Subbiah, J.~D. Kaplan, P.~Dhariwal,
  A.~Neelakantan, P.~Shyam, G.~Sastry, A.~Askell, S.~Agarwal, A.~Herbert-Voss,
  G.~Krueger, T.~Henighan, R.~Child, A.~Ramesh, D.~Ziegler, J.~Wu, C.~Winter,
  C.~Hesse, M.~Chen, E.~Sigler, M.~Litwin, S.~Gray, B.~Chess, J.~Clark,
  C.~Berner, S.~McCandlish, A.~Radford, I.~Sutskever, and D.~Amodei,
  ``{L}anguage {M}odels are {F}ew-{S}hot {L}earners,'' in \emph{Advances in
  Neural Information Processing Systems (NIPS)}.\hskip 1em plus 0.5em minus
  0.4em\relax Curran Associates, Inc., 2020, pp. 1877--1901. [Online].
  Available:
  \url{https://proceedings.neurips.cc/paper_files/paper/2020/file/1457c0d6bfcb4967418bfb8ac142f64a-Paper.pdf}
\BIBentrySTDinterwordspacing

\bibitem[Hadfield and Clark(2023)]{hadfield2023regulatory}
G.~K. Hadfield and J.~Clark, ``{R}egulatory {M}arkets: {T}he {F}uture of {AI}
  {G}overnance,'' \emph{arXiv preprint arXiv:2304.04914}, 2023.

\bibitem[Touvron et~al.(2023)Touvron, Martin, Stone, Albert, Almahairi, Babaei,
  Bashlykov, Batra, Bhargava, Bhosale, et~al.]{touvron2023llama}
H.~Touvron, L.~Martin, K.~Stone, P.~Albert, A.~Almahairi, Y.~Babaei,
  N.~Bashlykov, S.~Batra, P.~Bhargava, S.~Bhosale \emph{et~al.}, ``{L}lama 2:
  {O}pen foundation and fine-tuned chat models,'' \emph{arXiv preprint
  arXiv:2307.09288}, 2023.

\bibitem[Ferrandis(2022)]{ferrandis2022openrail}
\BIBentryALTinterwordspacing
C.~M. Ferrandis, ``{O}penrail: {T}owards open and responsible {AI} licensing
  frameworks,'' 2022. [Online]. Available: \url{https://www.licenses.ai/faq-2}
\BIBentrySTDinterwordspacing

\bibitem[Vincent(2023)]{llamaleak}
\BIBentryALTinterwordspacing
J.~Vincent, ``{M}eta’s powerful {AI} language model has leaked online —
  what happens now?'' 2023. [Online]. Available:
  \url{https://www.theverge.com/2023/3/8/23629362/meta-ai-language-model-llama-leak-online-misuse}
\BIBentrySTDinterwordspacing

\bibitem[Henderson et~al.(2023)Henderson, Mitchell, Manning, Jurafsky, and
  Finn]{henderson2023selfdestructing}
\BIBentryALTinterwordspacing
P.~Henderson, E.~Mitchell, C.~Manning, D.~Jurafsky, and C.~Finn,
  ``{S}elf-{D}estructing {M}odels: {I}ncreasing the {C}osts of {H}armful {D}ual
  {U}ses of {F}oundation {M}odels,'' in \emph{Proceedings of the 2023 AAAI/ACM
  Conference on AI, Ethics, and Society}, ser. AIES '23.\hskip 1em plus 0.5em
  minus 0.4em\relax Association for Computing Machinery, 2023, p. 287–296.
  [Online]. Available: \url{https://doi.org/10.1145/3600211.3604690}
\BIBentrySTDinterwordspacing

\bibitem[Lin et~al.(2024)Lin, Cui, Liao, and Wang]{lin2024malla}
Z.~Lin, J.~Cui, X.~Liao, and X.~Wang, ``{M}alla: {D}emystifying {R}eal-world
  {L}arge {L}anguage {M}odel {I}ntegrated {M}alicious {S}ervices,'' \emph{arXiv
  preprint arXiv:2401.03315}, 2024.

\bibitem[Choquette et~al.(2021)Choquette, Gandhi, Giroux, Stam, and
  Krashinsky]{choquette2021nvidia}
J.~Choquette, W.~Gandhi, O.~Giroux, N.~Stam, and R.~Krashinsky, ``{NVIDIA}
  {A}100 tensor core {GPU}: {P}erformance and innovation,'' \emph{IEEE Micro},
  no.~2, pp. 29--35, 2021.

\bibitem[Choquette(2023)]{choquette2023nvidia}
J.~Choquette, ``{NVIDIA} {H}opper {H}100 {GPU}: {S}caling {P}erformance,''
  \emph{IEEE Micro}, no.~3, pp. 9--17, 2023.

\bibitem[NVIDIA()]{NvidiaH200}
NVIDIA, ``{NVIDIA} {N}vidia {D}ata {C}enter {GPU} {R}esource {C}enter,''
  \url{https://resources.nvidia.com/l/en-us-gpu}, accessed: 2024-22-05.

\bibitem[Selig(2022)]{selig2022cerebras}
J.~Selig, ``{T}he cerebras software development kit: {A} technical overview,''
  2022.

\bibitem[Talpes et~al.(2022)Talpes, Williams, and
  Sarma]{talpes2023microarchitecture}
E.~Talpes, D.~Williams, and D.~D. Sarma, ``{DOJO}: {T}he {M}icroarchitecture of
  {T}esla’s {E}xa-{S}cale {C}omputer,'' in \emph{2022 IEEE Hot Chips 34
  Symposium (HCS)}, 2022, pp. 1--28.

\bibitem[Gwennap(2020)]{gwennap2020groq}
L.~Gwennap, ``{G}roq rocks neural networks,'' \emph{Microprocessor Report,
  Tech. Rep., jan}, 2020.

\bibitem[Chatha(2021)]{chatha2021qualcomm}
K.~Chatha, ``{Q}ualcomm{\textregistered} {C}loud {A}l 100: 12{TOPS}/{W}
  {S}calable, {H}igh {P}erformance and {L}ow {L}atency {D}eep {L}earning
  {I}nference {A}ccelerator,'' in \emph{2021 IEEE Hot Chips 33 Symposium
  (HCS)}.\hskip 1em plus 0.5em minus 0.4em\relax IEEE, 2021, pp. 1--19.

\bibitem[Jouppi et~al.(2021)Jouppi, Yoon, Ashcraft, Gottscho, Jablin, Kurian,
  Laudon, Li, Ma, Ma, et~al.]{jouppi2021ten}
N.~P. Jouppi, D.~H. Yoon, M.~Ashcraft, M.~Gottscho, T.~B. Jablin, G.~Kurian,
  J.~Laudon, S.~Li, P.~Ma, X.~Ma \emph{et~al.}, ``{T}en {L}essons from {T}hree
  {G}enerations {S}haped {G}oogle’s {TPU}v4i: {I}ndustrial {P}roduct,'' in
  \emph{2021 ACM/IEEE 48th Annual International Symposium on Computer
  Architecture (ISCA)}.\hskip 1em plus 0.5em minus 0.4em\relax IEEE, 2021, pp.
  1--14.

\bibitem[AMD()]{AMD}
``{AMD} {I}nstinct {MI}300 {S}eries {A}ccelerators,''
  \url{https://www.amd.com/en/products/accelerators/instinct/mi300.html},
  accessed: 2024-03-03.

\bibitem[Alok(2020)]{alok2020architecture}
G.~Alok, ``{A}rchitecture apocalypse dream architecture for deep learning
  inference and compute-versal ai core,'' \emph{Embedded World}, 2020.

\bibitem[Bursztein et~al.(2016)Bursztein, Malyshev, Pietraszek, and
  Thomas]{bursztein2016picasso}
E.~Bursztein, A.~Malyshev, T.~Pietraszek, and K.~Thomas, ``{P}icasso:
  {L}ightweight device class fingerprinting for web clients,'' in
  \emph{Proceedings of the 6th Workshop on Security and Privacy in Smartphones
  and Mobile Devices (SPSM)}, 2016, pp. 93--102.

\bibitem[Laor et~al.(2022)Laor, Mehanna, Durey, Dyadyuk, Laperdrix, Maurice,
  Oren, Rouvoy, Rudametkin, and Yarom]{Laor_2022}
\BIBentryALTinterwordspacing
T.~Laor, N.~Mehanna, A.~Durey, V.~Dyadyuk, P.~Laperdrix, C.~Maurice, Y.~Oren,
  R.~Rouvoy, W.~Rudametkin, and Y.~Yarom, ``{DRAWN} {APART}: {A} {D}evice
  {I}dentification {T}echnique based on {R}emote {GPU} {F}ingerprinting,'' in
  \emph{Proceedings 2022 Network and Distributed System Security Symposium
  (NDSS)}, ser. NDSS 2022.\hskip 1em plus 0.5em minus 0.4em\relax Internet
  Society, 2022. [Online]. Available:
  \url{http://dx.doi.org/10.14722/ndss.2022.24093}
\BIBentrySTDinterwordspacing

\bibitem[Outchakoucht et~al.(2017)Outchakoucht, Hamza, and
  Leroy]{outchakoucht2017dynamic}
A.~Outchakoucht, E.-S. Hamza, and J.~P. Leroy, ``Dynamic access control policy
  based on blockchain and machine learning for the internet of things,''
  \emph{International journal of advanced Computer Science and applications},
  vol.~8, no.~7, 2017.

\bibitem[Qi et~al.(2023)Qi, Zeng, Xie, Chen, Jia, Mittal, and
  Henderson]{qi2023fine}
X.~Qi, Y.~Zeng, T.~Xie, P.-Y. Chen, R.~Jia, P.~Mittal, and P.~Henderson,
  ``Fine-tuning aligned language models compromises safety, even when users do
  not intend to!'' \emph{arXiv preprint arXiv:2310.03693}, 2023.

\bibitem[Schl\"{o}gl et~al.(2023)Schl\"{o}gl, Hofer, and
  B\"{o}hme]{schlogl2024causes}
\BIBentryALTinterwordspacing
A.~Schl\"{o}gl, N.~Hofer, and R.~B\"{o}hme, ``{C}auses and {E}ffects of
  {U}nanticipated {N}umerical {D}eviations in {N}eural {N}etwork {I}nference
  {F}rameworks,'' in \emph{Advances in Neural Information Processing Systems
  (NIPS)}.\hskip 1em plus 0.5em minus 0.4em\relax Curran Associates, Inc.,
  2023, pp. 56\,095--56\,107. [Online]. Available:
  \url{https://proceedings.neurips.cc/paper_files/paper/2023/file/af076c3bdbf935b81d808e37c5ede463-Paper-Conference.pdf}
\BIBentrySTDinterwordspacing

\bibitem[Sabt et~al.(2015)Sabt, Achemlal, and Bouabdallah]{sabt2015trusted}
M.~Sabt, M.~Achemlal, and A.~Bouabdallah, ``Trusted execution environment: What
  it is, and what it is not,'' in \emph{2015 IEEE Trustcom/BigDataSE/Ispa},
  vol.~1.\hskip 1em plus 0.5em minus 0.4em\relax IEEE, 2015, pp. 57--64.

\bibitem[Goldreich(1998)]{goldreich1998secure}
O.~Goldreich, ``Secure multi-party computation,'' \emph{Manuscript. Preliminary
  version}, vol.~78, no. 110, pp. 1--108, 1998.

\bibitem[Naehrig et~al.(2011)Naehrig, Lauter, and
  Vaikuntanathan]{naehrig2011can}
M.~Naehrig, K.~Lauter, and V.~Vaikuntanathan, ``Can homomorphic encryption be
  practical?'' in \emph{Proceedings of the 3rd ACM workshop on Cloud computing
  security workshop}, 2011, pp. 113--124.

\bibitem[Mugunthan et~al.(2019)Mugunthan, Polychroniadou, Byrd, and
  Balch]{mugunthan2019smpai}
V.~Mugunthan, A.~Polychroniadou, D.~Byrd, and T.~H. Balch, ``Smpai: Secure
  multi-party computation for federated learning,'' in \emph{Proceedings of the
  NeurIPS 2019 Workshop on Robust AI in Financial Services}, vol.~21.\hskip 1em
  plus 0.5em minus 0.4em\relax MIT Press Cambridge, MA, USA, 2019.

\bibitem[Zhang et~al.(2020)Zhang, Li, Xia, Wang, Yan, and
  Liu]{zhang2020batchcrypt}
C.~Zhang, S.~Li, J.~Xia, W.~Wang, F.~Yan, and Y.~Liu, ``$\{$BatchCrypt$\}$:
  Efficient homomorphic encryption for $\{$Cross-Silo$\}$ federated learning,''
  in \emph{2020 USENIX annual technical conference (USENIX ATC 20)}, 2020, pp.
  493--506.

\bibitem[Chen et~al.(2020)Chen, Luo, Li, Xiang, Liu, and Li]{chen2020training}
Y.~Chen, F.~Luo, T.~Li, T.~Xiang, Z.~Liu, and J.~Li, ``A training-integrity
  privacy-preserving federated learning scheme with trusted execution
  environment,'' \emph{Information Sciences}, vol. 522, pp. 69--79, 2020.

\bibitem[Chakraborty et~al.(2020)Chakraborty, Mondai, and
  Srivastava]{chakraborty2020hardwareassisted}
A.~Chakraborty, A.~Mondai, and A.~Srivastava, ``Hardware-assisted intellectual
  property protection of deep learning models,'' in \emph{2020 57th ACM/IEEE
  Design Automation Conference (DAC)}, 2020, pp. 1--6.

\bibitem[Tram{\`e}r et~al.(2016)Tram{\`e}r, Zhang, Juels, Reiter, and
  Ristenpart]{tramer2016stealingapi}
\BIBentryALTinterwordspacing
F.~Tram{\`e}r, F.~Zhang, A.~Juels, M.~K. Reiter, and T.~Ristenpart,
  ``{S}tealing {M}achine {L}earning {M}odels via {P}rediction {API}s,'' in
  \emph{25th USENIX Security Symposium (USENIX Security 16)}.\hskip 1em plus
  0.5em minus 0.4em\relax USENIX Association, 2016, pp. 601--618. [Online].
  Available:
  \url{https://www.usenix.org/conference/usenixsecurity16/technical-sessions/presentation/tramer}
\BIBentrySTDinterwordspacing

\bibitem[Truong et~al.(2021)Truong, Maini, Walls, and Papernot]{truong2021data}
J.-B. Truong, P.~Maini, R.~J. Walls, and N.~Papernot, ``{D}ata-{F}ree {M}odel
  {E}xtraction,'' in \emph{Proceedings of the IEEE/CVF conference on computer
  vision and pattern recognition (CVPR)}, 2021, pp. 4771--4780.

\bibitem[Shafran et~al.(2023)Shafran, Shumailov, Erdogdu, and
  Papernot]{shafran2023labeling}
A.~Shafran, I.~Shumailov, M.~A. Erdogdu, and N.~Papernot, ``{B}eyond {L}abeling
  {O}racles: {W}hat does it mean to steal {ML} models?'' 2023.

\bibitem[Ma et~al.(2023)Ma, Qiu, Gao, Zhang, Abuadbba, Xue, Fu, Zhang,
  Al-Sarawi, and Abbott]{ma2023quantization}
H.~Ma, H.~Qiu, Y.~Gao, Z.~Zhang, A.~Abuadbba, M.~Xue, A.~Fu, J.~Zhang, S.~F.
  Al-Sarawi, and D.~Abbott, ``{Q}uantization {B}ackdoors to {D}eep {L}earning
  {C}ommercial {F}rameworks,'' \emph{IEEE Transactions on Dependable and Secure
  Computing}, 2023.

\bibitem[Hong et~al.(2021)Hong, Panaitescu-Liess, Kaya, and
  Dumitras]{hong2021quantization}
\BIBentryALTinterwordspacing
S.~Hong, M.-A. Panaitescu-Liess, Y.~Kaya, and T.~Dumitras,
  ``{Q}u-{ANTI}-zation: {E}xploiting {Q}uantization {A}rtifacts for {A}chieving
  {A}dversarial {O}utcomes,'' in \emph{Advances in Neural Information
  Processing Systems (NIPS)}.\hskip 1em plus 0.5em minus 0.4em\relax Curran
  Associates, Inc., 2021, pp. 9303--9316. [Online]. Available:
  \url{https://proceedings.neurips.cc/paper_files/paper/2021/file/4d8bd3f7351f4fee76ba17594f070ddd-Paper.pdf}
\BIBentrySTDinterwordspacing

\bibitem[LeCun et~al.(1989)LeCun, Denker, and Solla]{lecun1989optimal}
\BIBentryALTinterwordspacing
Y.~LeCun, J.~Denker, and S.~Solla, ``{O}ptimal {B}rain {D}amage,'' in
  \emph{Advances in Neural Information Processing Systems (NIPS)}.\hskip 1em
  plus 0.5em minus 0.4em\relax Morgan-Kaufmann, 1989. [Online]. Available:
  \url{https://proceedings.neurips.cc/paper_files/paper/1989/file/6c9882bbac1c7093bd25041881277658-Paper.pdf}
\BIBentrySTDinterwordspacing

\bibitem[Van~Aubel et~al.(2015)Van~Aubel, Bernstein, and
  Niederhagen]{vanaubel2015investigating}
P.~Van~Aubel, D.~J. Bernstein, and R.~Niederhagen, ``{I}nvestigating {SRAM}
  {PUF}s in large {CPU}s and {GPU}s,'' in \emph{Security, Privacy, and Applied
  Cryptography Engineering}.\hskip 1em plus 0.5em minus 0.4em\relax Springer
  International Publishing, 2015, pp. 228--247.

\bibitem[Li et~al.(2015)Li, Fu, and Luo]{fengjun2015freqpuf}
\BIBentryALTinterwordspacing
F.~Li, X.~Fu, and B.~Luo, ``{POSTER}: {A} {H}ardware {F}ingerprint {U}sing
  {GPU} core frequency variations,'' in \emph{Proceedings of the 22nd ACM
  SIGSAC Conference on Computer and Communications Security}, ser. CCS
  '15.\hskip 1em plus 0.5em minus 0.4em\relax Association for Computing
  Machinery, 2015, p. 1650–1652. [Online]. Available:
  \url{https://doi.org/10.1145/2810103.2810105}
\BIBentrySTDinterwordspacing

\bibitem[Forlin et~al.(2020)Forlin, Husemann, Carro, Reinbrecht, Hamdioui, and
  Taouil]{forlin2020g}
B.~Forlin, R.~Husemann, L.~Carro, C.~Reinbrecht, S.~Hamdioui, and M.~Taouil,
  ``{G-PUF}: {A}n {I}ntrinsic {PUF} {B}ased on {GPU} {E}rror {S}ignatures,'' in
  \emph{2020 IEEE European Test Symposium (ETS)}, 2020, pp. 1--2.

\bibitem[Dodis et~al.(2004)Dodis, Reyzin, and Smith]{Dodis_2008}
Y.~Dodis, L.~Reyzin, and A.~Smith, ``{F}uzzy {E}xtractors: {H}ow to {G}enerate
  {S}trong {K}eys from {B}iometrics and {O}ther {N}oisy {D}ata,'' in
  \emph{Advances in Cryptology - EUROCRYPT 2004}.\hskip 1em plus 0.5em minus
  0.4em\relax Springer Berlin Heidelberg, 2004, pp. 523--540.

\bibitem[Kelsey et~al.(1997)Kelsey, Schneier, Hall, and
  Wagner]{kelsey1997secure}
J.~Kelsey, B.~Schneier, C.~Hall, and D.~Wagner, ``Secure applications of
  low-entropy keys,'' in \emph{International Workshop on Information
  Security}.\hskip 1em plus 0.5em minus 0.4em\relax Springer, 1997, pp.
  121--134.

\bibitem[Li et~al.(2017)Li, Kadav, Durdanovic, Samet, and Graf]{li2016pruning}
\BIBentryALTinterwordspacing
H.~Li, A.~Kadav, I.~Durdanovic, H.~Samet, and H.~P. Graf, ``{P}runing {F}ilters
  for {E}fficient {C}onv{N}ets,'' in \emph{5th International Conference on
  Learning Representations (ICLR)}.\hskip 1em plus 0.5em minus 0.4em\relax
  OpenReview.net, 2017. [Online]. Available:
  \url{https://openreview.net/forum?id=rJqFGTslg}
\BIBentrySTDinterwordspacing

\bibitem[Devlin et~al.(2019)Devlin, Chang, Lee, and Toutanova]{devlin2018bert}
\BIBentryALTinterwordspacing
J.~Devlin, M.~Chang, K.~Lee, and K.~Toutanova, ``{BERT:} {P}re-training of
  {D}eep {B}idirectional {T}ransformers for {L}anguage {U}nderstanding,'' in
  \emph{Proceedings of the 2019 Conference of the North American Chapter of the
  Association for Computational Linguistics: Human Language Technologies
  (NAACL-HLT)}.\hskip 1em plus 0.5em minus 0.4em\relax Association for
  Computational Linguistics, 2019, pp. 4171--4186. [Online]. Available:
  \url{https://doi.org/10.18653/v1/n19-1423}
\BIBentrySTDinterwordspacing

\bibitem[Wang et~al.(2018)Wang, Singh, Michael, Hill, Levy, and
  Bowman]{wang2018glue}
\BIBentryALTinterwordspacing
A.~Wang, A.~Singh, J.~Michael, F.~Hill, O.~Levy, and S.~Bowman, ``{GLUE}: {A}
  {M}ulti-{T}ask {B}enchmark and {A}nalysis {P}latform for {N}atural {L}anguage
  {U}nderstanding,'' in \emph{Proceedings of the 2018 {EMNLP} Workshop
  {B}lackbox{NLP}: Analyzing and Interpreting Neural Networks for {NLP}}.\hskip
  1em plus 0.5em minus 0.4em\relax Association for Computational Linguistics,
  2018, pp. 353--355. [Online]. Available:
  \url{https://aclanthology.org/W18-5446}
\BIBentrySTDinterwordspacing

\bibitem[Krizhevsky and Hinton(2009)]{krizhevsky2014cifar}
\BIBentryALTinterwordspacing
A.~Krizhevsky and G.~Hinton, ``{L}earning {M}ultiple {L}ayers of {F}eatures
  from {T}iny {I}mages,'' University of Toronto, Tech. Rep., 2009. [Online].
  Available:
  \url{https://www.cs.toronto.edu/~kriz/learning-features-2009-TR.pdf}
\BIBentrySTDinterwordspacing

\bibitem[Nilsback and Zisserman(2008)]{nilsback2008automated}
M.-E. Nilsback and A.~Zisserman, ``{A}utomated flower classification over a
  large number of classes,'' in \emph{2008 Sixth Indian conference on computer
  vision, graphics \& image processing}.\hskip 1em plus 0.5em minus 0.4em\relax
  IEEE, 2008, pp. 722--729.

\bibitem[He et~al.(2016)He, Zhang, Ren, and Sun]{he2016deep}
K.~He, X.~Zhang, S.~Ren, and J.~Sun, ``{D}eep {R}esidual {L}earning for {I}mage
  {R}ecognition,'' in \emph{Proceedings of the IEEE conference on computer
  vision and pattern recognition (CVPR)}, 2016, pp. 770--778.

\bibitem[Dosovitskiy et~al.(2021)Dosovitskiy, Beyer, Kolesnikov, Weissenborn,
  Zhai, Unterthiner, Dehghani, Minderer, Heigold, Gelly, Uszkoreit, and
  Houlsby]{dosovitskiy2020image}
\BIBentryALTinterwordspacing
A.~Dosovitskiy, L.~Beyer, A.~Kolesnikov, D.~Weissenborn, X.~Zhai,
  T.~Unterthiner, M.~Dehghani, M.~Minderer, G.~Heigold, S.~Gelly, J.~Uszkoreit,
  and N.~Houlsby, ``{A}n {I}mage is {W}orth 16x16 {W}ords: {T}ransformers for
  {I}mage {R}ecognition at {S}cale,'' in \emph{9th International Conference on
  Learning Representations (ICLR)}.\hskip 1em plus 0.5em minus 0.4em\relax
  OpenReview.net, 2021. [Online]. Available:
  \url{https://openreview.net/forum?id=YicbFdNTTy}
\BIBentrySTDinterwordspacing

\bibitem[Cheng et~al.(2023)Cheng, Zhang, Yu, Montgomerie-Corcoran, Xiao,
  Bouganis, and Zhao]{cheng2023fast}
J.~Cheng, C.~Zhang, Z.~Yu, A.~Montgomerie-Corcoran, C.~Xiao, C.-S. Bouganis,
  and Y.~Zhao, ``{F}ast {P}rototyping {N}ext-{G}eneration {A}ccelerators for
  {N}ew {ML} {M}odels using {MASE}: {ML} {A}ccelerator {S}ystem
  {E}xploration,'' \emph{arXiv preprint arXiv:2307.15517}, 2023.

\end{thebibliography}
\FloatBarrier

\appendix

\subsection{Broader Impact}
\label{appendix:impact}
Our research addresses the growing need to protect machine learning models from misuse. By introducing the concept of ML Hardware Locking, our work offers a new tool to safeguard machine learning IP and aids responsible ML development and deployment. Our work also has implications for the governance of ML. By connecting models to specific hardware, another tool becomes available to control where and how such models are used. This could be particularly valuable in safety-critical applications, where ensuring that models are only executed in authenticated settings is paramount.

We recognize the potential for ML Hardware Locking to impact access and fairness in the AI landscape. Tying models to specific hardware could inadvertently create barriers for individuals or organisations with limited resources or access to authorized hardware. At the same time, ML Hardware Locking also provides a conceptually new way to enable offline access to models that previously could not be accessed at all.

\subsection{Experiment Compute Resources}
\label{appendix:experiment-resources}

The soft locking experiments consume most of the compute resources. We conducted all sparsity-aware locking and quantization-aware locking on NVIDIA V100 GPUs and 18-core Intel Xeon (Broadwell) processors. The fine-tuning took around 1 GPU hours per trial on average, and in total, the fine-tuning time was around 180 GPU hours. We spent additional time on preliminary and failed experiments, which is around 40 GPU hours in total. The emulation cost experiments were performed on three NVIDIA RTXA6000 GPUs with an AMD EPYC 7713 64-core processor. The emulation cost experiments took around 4 GPU hours. The hard locking experiments were conducted on NVIDIA GTX1080Ti, RTX2080Ti, RTX3090, RTXA6000 GPUs, and took 10 GPU hours in total.

\subsection{Formal Definition of Locking Mechanisms}\label{appendix:definitions}

This paper aims to develop locking mechanisms $L : M \times H \rightarrow M$ which make it difficult to use a machine learning model $m \in M$ on an unauthorized hardware platform $u \in U$ whilst not significantly affecting the behaviour of $m$ on authorized hardware platforms $a \in A$, where $A \cup U = H$ is the set of all hardware.

The performance of the model $m$ on hardware $h$ is given by $P (m, h) \in [0, 1]$, where higher values indicate better performance and $0$ corresponds to random guessing. Similarly, the time taken by a model $m$ to run on hardware $h$ is given by $T (m, h) \in \mathcal{R}^+$ .

We use these metrics to define two types of locked models: soft locked models and hard locked models.

\subsubsection{Definition: soft locked model}

A soft locked model $L_{soft} (m, A)$ on authorized hardware behaves the same as the original model $m$, but performs significantly worse or is significantly slower on unauthorized hardware:

\begin{align*}
	(\forall a \in A. P(L_{soft} (m, A), a) \approx P(m, a) \\
	\land T(L_{soft} (m, A), a) \approx T (m, a)) \\
	\land (\forall u \in U. P (L_{soft} (m, A), u) \ll P (m, u) \\
	\vee T (L_{soft} (m, A), u) \gg T (m, u))
\end{align*}

\subsubsection{Definition: hard locked model}

A hard locked model $L_{hard} (m, A)$ on authorized hardware behaves the same as the original model $m$, but is no better than random guessing on unauthorized hardware:

\begin{align*}
	(\forall a \in A. P (L_{hard}(m, A), a) \approx P (m, a)\\
	\wedge T(L_{hard} (m, a), a) \approx T (m, a)) \\
	\wedge (\forall u \in U. P (L_{hard} (m, A), u) \approx P (m_{rand}, u))
\end{align*}

\subsection{Hard Locking and Entropy Estimates}\label{sec:hard-locking-entropy}

Results are presented in~\Cref{tab:fingerprints}. Here, we list 20 as a strict upper bound for the clock fingerprint as it is the number of bits in the output. We expect the real entropy to still be considerable to provide useful security guarantees, but cannot estimate it without a large-scale experiment on diverse hardware.

Twice the \textrm{number of convolutions} is a theoretical result based on \citet{schlogl2024causes}'s work which finds four equivalence classes for each convolution performed. It is an upper bound for a convolution-based finite precision fingerprint, but convolution is not the only possible strategy for a finite precision fingerprint, indeed our finite precision fingerprints in~\Cref{sec:finite_precision_code} and \autoref{sec:finite-precision-fingerprint-data} are not based on convolutions.
The entropy of the SRAM PUF is an experimental result \citep{vanaubel2015investigating}, theoretically more than 256 is possible, but in our implementation encryption schemes we assume 256 bits, so any further entropy is not applicable.

\begin{table}[H]
\caption{Comparison of methods of generating fingerprints from hardware. * estimated}
\centering
\begin{tabular}{llr}
\toprule

  \textbf{Method}
& \textbf{Entropy (bits)}
& \textbf{Error rate} \\

\midrule

Clock                                       & 20 (upper bound)    & <0.1\%  \\
Finite precision                            & 2$\times \textrm{num. of}$ & 5\%*  \\
                                            & convolutions (theoretical) & \\
SRAM PUF \citep{vanaubel2015investigating}  & >256  & 5\%*    \\

\bottomrule
\end{tabular}

\label{tab:fingerprints}

\end{table}

\newpage
\subsection{Base-Model Training}

\subsubsection{Sparsity-aware lock} We use open-source SoTA setups for training the base models in soft locking experiments:

\begin{itemize}
    \item We use an open-source Github checkpoint\footnote{Github repository: \href{https://github.com/huyvnphan/PyTorch_CIFAR10}{huyvnphan/PyTorch\_CIFAR10}} as the base model for \textit{ResNet18/50} on CIFAR10. For \textit{ResNet18/50} on CIFAR100 and Flowers102, we train the models from scratch using the open-source scripts\footnote{Github repository: \href{https://github.com/weiaicunzai/pytorch-cifar100}{weiaicunzai/pytorch-cifar100}}. We adopt their hyperparameter settings except that we resize the Flowers102 images to 128$\times$128.
    \item We download \textit{ViT\_B\_16-224} checkpoint from an open-sourced Github repository\footnote{Github repository: \href{https://github.com/jeonsworld/ViT-pytorch}{jeonsworld/ViT-pytorch}} and fine-tune it using the scripts and hyperparameters provided in the same repository. All images are resised to 224$\times$224 during training.
    \item We download \textit{bert-base-cased} checkpoint from HuggingFace\footnote{HuggingFace checkpoint: \href{https://huggingface.co/google-bert/bert-base-cased}{google-bert/bert-base-cased}}, and use the default hyperparameters provided in the sequence classification training script open-sourced in the transformers repository\footnote{HuggingFace transformer: \href{https://github.com/huggingface/transformers/blob/main/examples/pytorch/text-classification/run_glue.py}{run\_glue.py} for sequence classification}.
\end{itemize}

\subsubsection{Quantisation-aware lock} We use the same checkpoints and settings for quantization-aware locking experiments. The checkpoints were trained for 50 epochs with a quantization-aware loss using AdamW with a learning-rate of $1e^{-3}$, a batch size of 256, and a random seed of $0$.

\subsection{Soft-Locking Fine-Tuning}
\label{hparams}
We use AdamW with a learning-rate of $1e^{-5}$ for both soft-locking procedures. \textit{ViT-B\_16-224} and the \textit{ResNet} models were locked with batch-sizes of 32 and 256 respectively. Flowers102 images were resised to $128\times128$ for the ResNet models, and all datasets were resized to $224\times224$ for \textit{ViT-B\_16-224}. All runs were seed with random seed $0$. \\

\subsection{Sparsity-Aware Locking - Extended \Cref{tab:pruning}}
We present below a finer version of~\Cref{tab:pruning} below, with two extra sparsity levels, $p = 0.10, 0.25$, for completeness.

\begin{table}[H]
\caption{Results are presented as $\mathit{Acc}_{\text{authorized}}^{\text{locked}} (\textcolor{czgreen}{\Delta_\text{orig}}, \textcolor{czred}{\Delta_{\text{lock}}})$}
\centering
\adjustbox{max width=\linewidth}{
\begin{tabular}{lccccc}
\toprule
 & \multicolumn{5}{c}{\textbf{Pruning Levels}} \\
 \textbf{Dataset} & 0.05 & 0.10 & 0.25 & 0.50 & 0.75 \\
\midrule
 & \multicolumn{5}{c}{\textit{BERT}} \\
CoLA & 0.80 (-0.03, 0.49) & 0.83 (-0.01, 0.52) & 0.84 (0.00, 0.53) & 0.83 (0.00, 0.53) & 0.84 (0.00, 0.53) \\
MRPC & 0.84 (-0.02, 0.15) & 0.84 (-0.02, 0.16) & 0.86 (0.00, 0.18) & 0.86 (0.00, 0.54) & 0.87 (0.00, 0.55) \\
SST-2 & 0.92 (-0.01, 0.43) & 0.92 (0.00, 0.43) & 0.93 (0.00, 0.43) & 0.92 (0.00, 0.43) & 0.92 (0.00, 0.43) \\
\midrule
 & \multicolumn{5}{c}{\textit{Resnet18}} \\
CIFAR10 & 0.89 (0.03, 0.67) & 0.92 (0.00, 0.78) & 0.93 (0.00, 0.83) & 0.93 (0.00, 0.83) & 0.93 (0.00, 0.83) \\
CIFAR100 & 0.73 (0.03, 0.71) & 0.75 (0.01, 0.74) & 0.76 (0.00, 0.75) & 0.76 (0.00, 0.75) & 0.76 (0.13, 0.75) \\
Flowers102 & 0.84 (0.04, 0.83) & 0.84 (0.04, 0.84) & 0.89 (0.00, 0.88) & 0.89 (0.00, 0.88) & 0.89 (0.00, 0.88) \\
\midrule
 & \multicolumn{5}{c}{\textit{Resnet50}} \\
CIFAR10 & 0.92 (0.01, 0.81) & 0.93 (0.00, 0.83) & 0.93 (-0.01, 0.83) & 0.94 (-0.01, 0.84) & 0.94 (-0.01, 0.84) \\
CIFAR100 & 0.69 (0.09, 0.63) & 0.76 (0.02, 0.75) & 0.78 (0.00, 0.77) & 0.78 (0.00, 0.77) & 0.78 (0.00, 0.77) \\
Flowers102 & 0.76 (0.10, 0.74) & 0.82 (0.04, 0.80) & 0.86 (0.00, 0.84) & 0.86 (0.00, 0.84) & 0.85 (0.00, 0.84) \\
\midrule
 & \multicolumn{5}{c}{\textit{ViT-B\_16-224}} \\
CIFAR10 & 0.99 (0.00, 0.89) & 0.99 (0.00, 0.89) & 0.99 (0.00, 0.89) & 0.99 (0.07, 0.89) & 0.93 (-0.02, 0.93) \\
CIFAR100 & 0.92 (-0.02, 0.92) & 0.93 (-0.02, 0.92) & 0.93 (-0.02, 0.92) & 0.93 (-0.02, 0.93) & 0.93 (-0.02, 0.93) \\
Flowers102 & 1.00 (0.00, 0.98) & 1.00 (0.00, 0.98) & 1.00 (0.00, 0.99) & 1.00 (0.00, 0.99) & 1.00 (0.00, 1.00) \\
\bottomrule
\end{tabular}}
\label{tab:pruning_extended}
\end{table}

\subsection{Baseline Performance Degradation}
\label{soft-base}
\subsubsection{Sparsity-aware lock}
We present below the degradation in performance from the transfer of the dense \textit{original} model, i.e. before locking, to a higher-levels of sparsity, $p$.

\begin{table}[H]
\caption{Results are presented as $Acc_{\text{authorized}}(\Delta_{\text{base}} = Acc_{\text{authorized}} - Acc_{\text{un\-authorized}})$}
\centering
\adjustbox{max width=\linewidth}{
\begin{tabular}{lccccc}
\toprule
& \multicolumn{5}{c}{\textbf{Pruning Levels}} \ \\
\textbf{Dataset} & 0.05 & 0.10 & 0.25 & 0.50 & 0.75 \ \\
\midrule
& \multicolumn{5}{c}{\textit{BERT}} \ \\
MRPC & 0.86 (0.00) & 0.86 (0.00) & 0.86 (0.01) & 0.86 (0.18) & 0.86 (0.54) \ \\
SST-2 & 0.92 (-0.00) & 0.92 (-0.00) & 0.92 (-0.00) & 0.92 (0.03) & 0.92 (0.30) \ \\
CoLA & 0.84 (-0.00) & 0.84 (-0.00) & 0.84 (0.01) & 0.84 (0.53) & 0.84 (0.53) \ \\
\midrule
& \multicolumn{5}{c}{\textit{Resnet18}} \ \\
CIFAR10 & 0.93 (0.00) & 0.93 (0.00) & 0.93 (0.00) & 0.93 (0.06) & 0.93 (0.68) \ \\
CIFAR100 & 0.76 (0.00) & 0.76 (-0.00) & 0.76 (0.01) & 0.76 (0.07) & 0.76 (0.08) \ \\
Flowers102 & 0.89 (0.00) & 0.89 (0.01) & 0.89 (0.01) & 0.89 (0.08) & 0.89 (0.68) \ \\
\midrule
& \multicolumn{5}{c}{\textit{Resnet50}} \ \\
CIFAR10 & 0.93 (-0.00) & 0.93 (-0.00) & 0.93 (0.00) & 0.93 (0.06) & 0.93 (0.70) \ \\
CIFAR100 & 0.78 (-0.00) & 0.78 (-0.00) & 0.78 (0.01) & 0.78 (0.07) & 0.78 (0.65) \ \\
Flowers102 & 0.86 (0.00) & 0.86 (0.00) & 0.86 (0.01) & 0.86 (0.10) & 0.86 (0.66) \ \\
\midrule
& \multicolumn{5}{c}{\textit{ViT-B\_16-224}} \ \\
CIFAR10 & 0.99 (-0.00) & 0.99 (-0.00) & 0.99 (0.00) & 0.99 (0.07) & 0.99 (0.00) \ \\
CIFAR100 & 0.91 (0.00) & 0.91 (0.01) & 0.91 (0.01) & 0.91 (0.30) & 0.91 (0.90) \ \\
Flowers102 & 1.00 (0.00) & 1.00 (0.00) & 1.00 (0.00) & 1.00 (0.23) & 1.00 (0.98) \ \\
\bottomrule
\end{tabular}}
\label{tab:pruning_appendix_baseline}
\end{table}

As seen from \Cref{tab:pruning_appendix_baseline}, in most settings, models can consistently be deployed at a pruning level of up to $p = 0.5$, without incurring a notable cost in terms of accuracy. In certain cases, for example \textit{ResNet18} on CIFAR100 and \textit{ViT-B\_16-224} on CIFAR10, even a sparsity level of $p = 0.75$ does not significantly affect performance. \Cref{tab:pruning_appendix_baseline} therefore highlights that the significant degradation in performance seen when deploying locked models on unauthorized levels of sparsity \ref{tab:pruning} can attributed to sparsity-aware locking, and not merely pruning.

\subsubsection{Quantisation-aware lock}
We present below the degradation in performance from the transfer of the \textit{original} model, i.e. before locking, to different arithmetics.

\begin{table}[H]
\caption{Results are presented as $Acc_{\text{authorized}}(\Delta_{\text{base}} = Acc_{\text{authorized}} - Acc_{\text{un-authorized}})$}
\centering
\begin{tabular}{llll}
\toprule
 authorized $\rightarrow$ Unauthorized & \textbf{Resnet18} & \textbf{Resnet50} \\
\midrule
FP32 $\rightarrow$ 8-bit MiniFloat
& 0.88 (0.00) & 0.88 (0.01) \\
Int8 $\rightarrow$ 8-bit MiniFloat & 0.88 (0.00) & 0.88 (0.01) \\
16-bit MiniFloat $\rightarrow$ Int8 & 0.88 (0.00) & 0.88 (0.01)\\
FP16 $\rightarrow$ Int8 & 0.88 (0.00) & 0.88 (-0.01)\\
\bottomrule
\end{tabular}
\label{tab:quan_appendix}
\end{table}

There are negligible performance drops, if any, from quantizing the base model to an un-authorized arithmetic, as seen by the low $\Delta_{\text{base}}$ values. Note that this is as the base-model was partially trained with a quantization-aware loss at \textbf{Int8}. Like with sparsity-aware locking, Table \ref{tab:quan_appendix} clearly shows that the degradation in accuracy when the locked model is transferred to unauthorized arithmetic schemes \ref{tab:quan} is a result of soft-locking, and not merely from the differences in model representation across hardware.

\subsection{Soft-Locking Optimisation Profiles}
\label{appendix:optimisation_profiles}

\subsubsection{Sparsity-aware lock}
We present below the validation curves of the soft-locking optimisation procedure for the three vision models for a subset of the datasets, \textit{CIFAR10, CIFAR100}, and pruning proportions, $p=\{0.05, 0.25, 0.50\}$.
\begin{figure}[H]
    \centering
    \includegraphics[width=\linewidth]{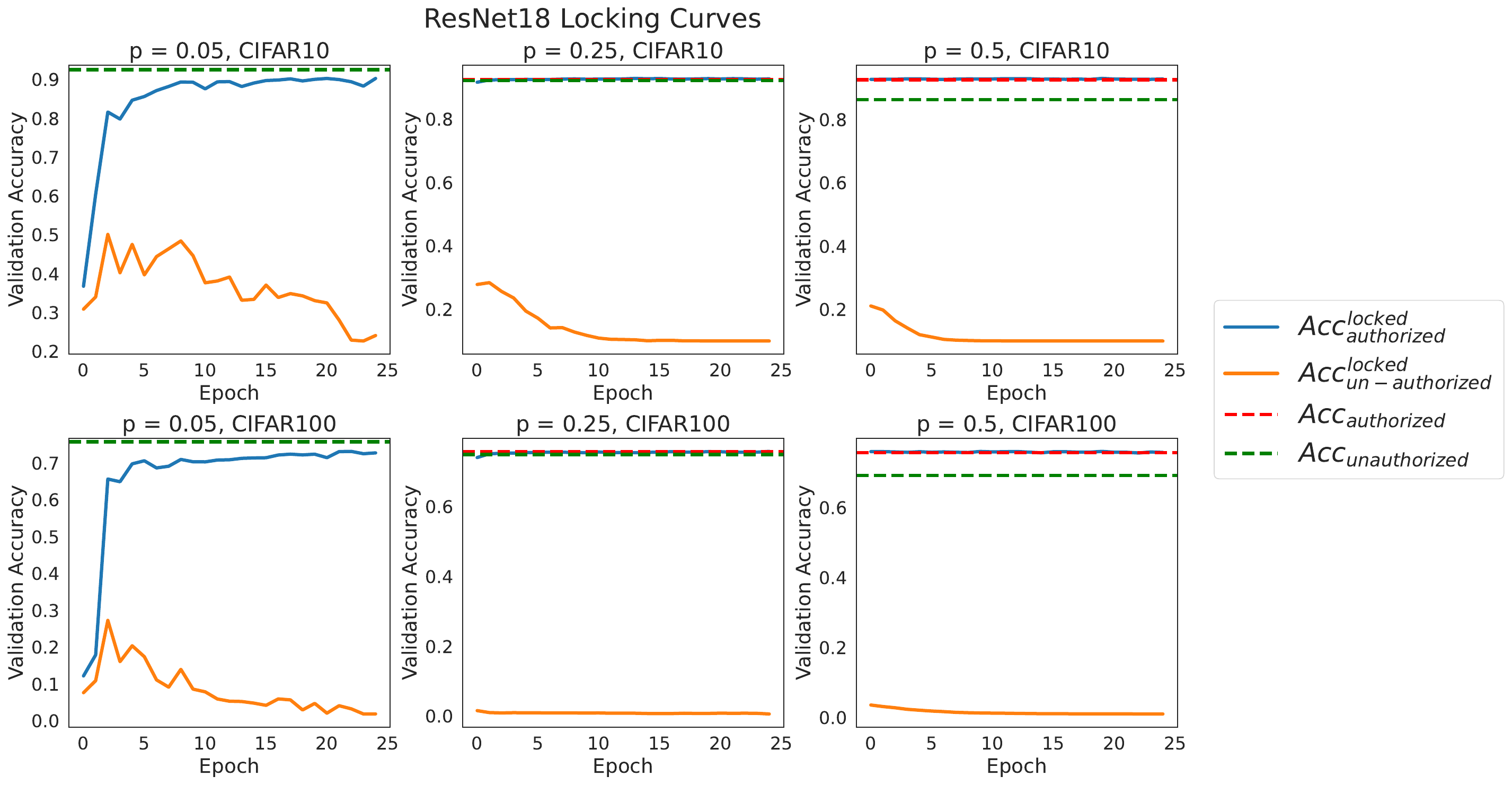}
    \caption{\textit{ResNet18} locking curves}
    \label{fig:resnet18-soft-prune}
\end{figure}
\begin{figure}[H]
    \centering
    \includegraphics[width=\linewidth]{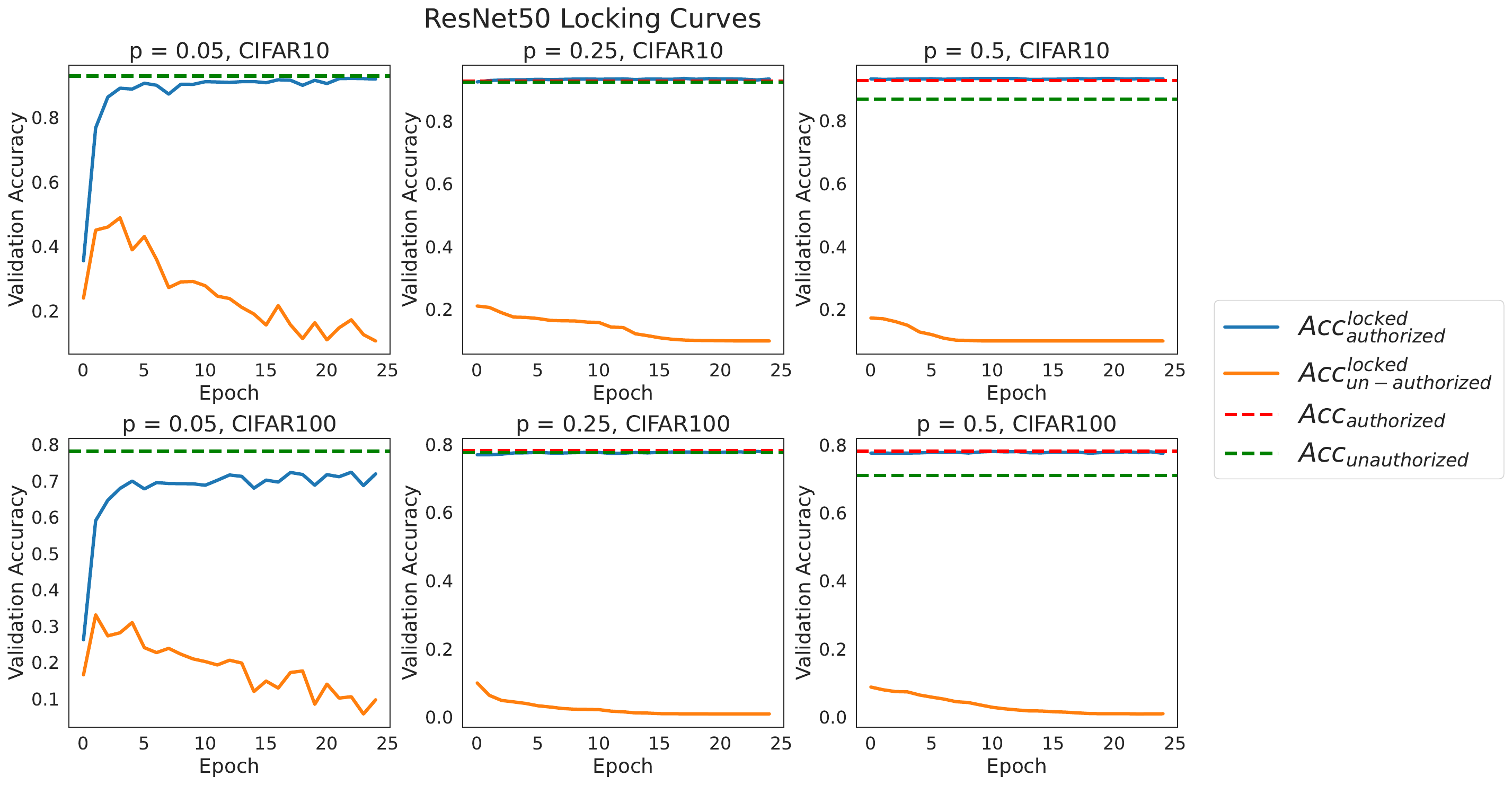}
    \caption{\textit{ResNet50} locking curves}
    \label{fig:resnet50-soft-prune}
\end{figure}
\begin{figure}[H]
    \centering
    \includegraphics[width=\linewidth]{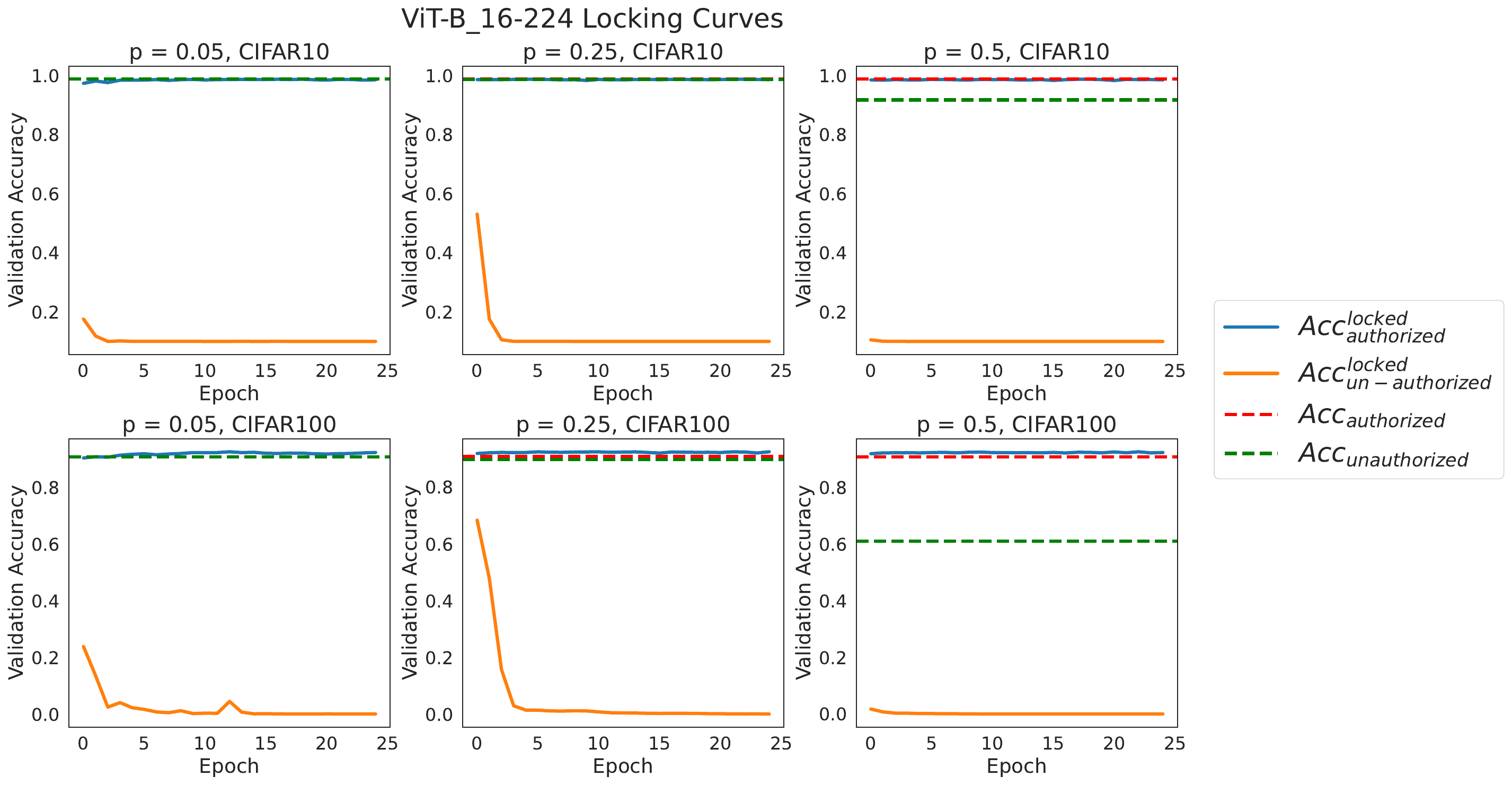}
    \caption{\textit{ViT-B\_16-224} locking curves}
    \label{fig:vit-soft-prune}
\end{figure}
The optimisation profiles generally converge much more rapidly, across datasets and models, for higher pruning proportions, $p$. This reinforces the intuitive result that the optimisation solved is \textit{easier} for greater sparsity values, and hence, larger values of $p$. For $p = 0.05$, the performance of the authorized locked model ($Acc_{authorized}^{locked}$) first reduces in performance, suggesting the optimisation is dominated by the second term of the manipulation loss function. Across the course of the optimisation, this accuracy increases to an acceptable level, due to the eventual effect of the first term of the loss function.

The notable exception to this trend is \textit{ViT-B\_16-224}. We posit that this is due to its increased size (in number of parameters), which allows for a decoupling of the opposing optimisation objectives, as the number of parameters at $5\%$ is much greater in \textit{ViT-B\_16-224} than in the \textit{ResNet} models. The larger number of parameters can be used to effectively encode the differential knowledge between the authorized and unauthorized variants of the model.

\subsubsection{Quantisation-aware lock}
We present below the validation curves of the quantization-aware locking procedure for the ResNet models on \textit{CIFAR10, CIFAR100} - for the authorized, unauthorized arithmetic pairs presented in \ref{tab:quan}. Note the failure of the locking optimisation procedure for \textbf{MiniFloat8} $\rightarrow$ \textbf{Int8}.

\begin{figure}[H]
    \centering
    \includegraphics[width=\linewidth]{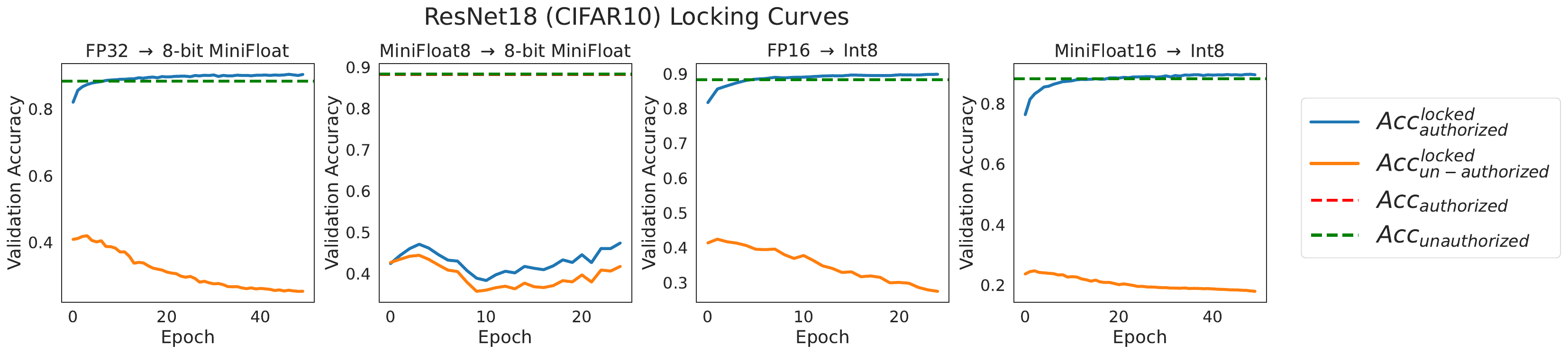}
    \caption{\textit{ResNet18} locking curves}
    \label{fig:resnet50-quant}
\end{figure}
\begin{figure}[H]
    \centering
    \includegraphics[width=\linewidth]{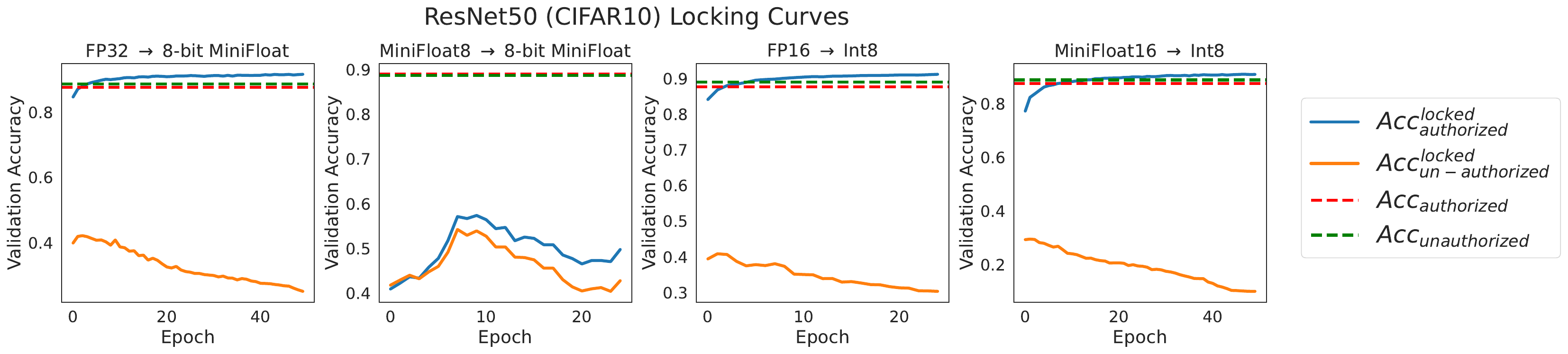}
    \caption{\textit{ResNet50} locking curves}
    \label{fig:resnet50-quant}
\end{figure}

\newpage
\subsection{Sparsity-Aware Locking}
\label{appendix:recovery}
In the following subsection, we present various miscellaneous findings regarding the sparsity-aware lock.

\subsubsection{Specificity of sparsity-aware locking}
 We present below the the performance of locked model across a range of sparsity levels for the \textit{ResNet} models on CIFAR10.

\begin{figure}[H]
    \centering
    \includegraphics[width=\linewidth]{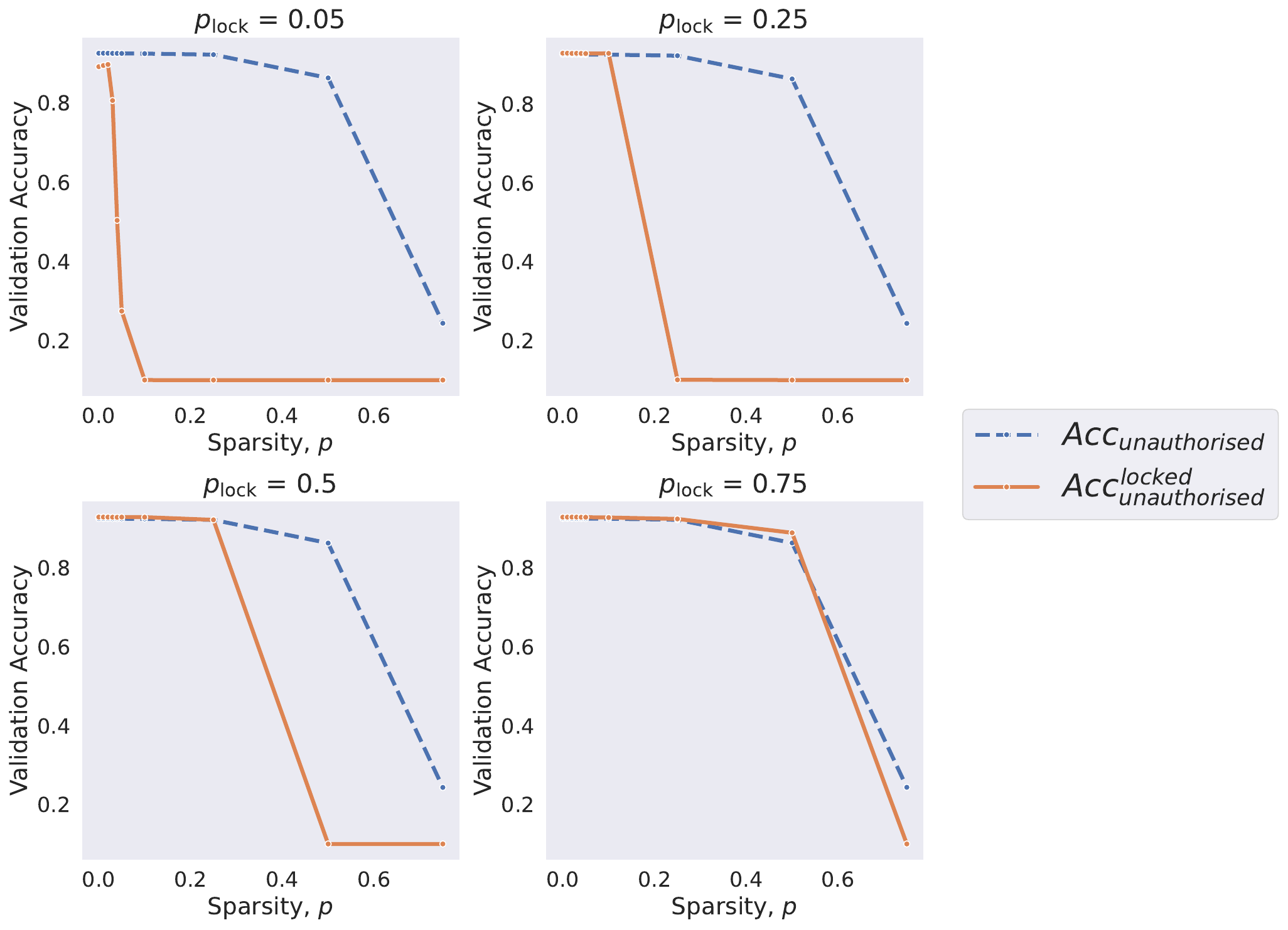}
    \caption{Validation accuracy across sparsity levels (\textit{ResNet18} on  \textit{CIFAR10})}
    \label{fig:specificity_cifar10_18}
\end{figure}

\begin{figure}[H]
    \centering
    \includegraphics[width=\linewidth]{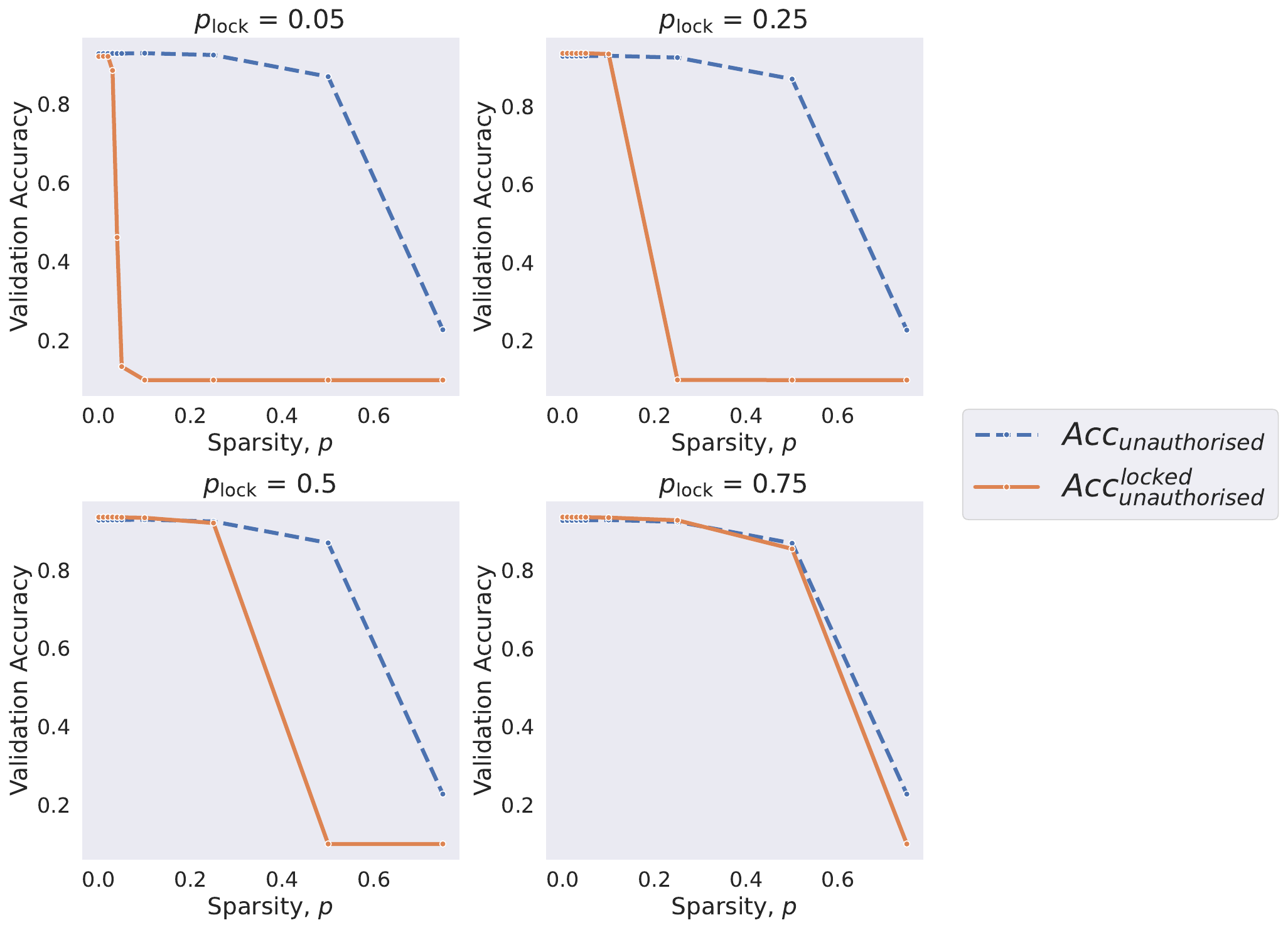}
    \caption{Validation accuracy across sparsity levels (\textit{ResNet50} on  \textit{CIFAR10})}
    \label{fig:specificity_cifar10_50}
\end{figure}

In both models and all locking levels, $p_\text{lock}$, there is a highly localised drop in accuracy at the sparsity-level that is locked. This is desirable for setting in which the provider that is locking a proprietary model is aware of the downstream sparsity-level utilised by unauthorized users. If this is not known, or there is no single sparsity level that is unauthorized, models can be conservatively locked at $p_\text{lock}$ = 0.05. As seen in \ref{fig:specificity_cifar10_18} and \ref{fig:specificity_cifar10_50}, locking the model at $p_\text{lock}$ = 0.05 effectively renders transfer to any level of sparsity that enables efficient execution on hardware useless.

\subsubsection{Effect of sparsity-aware locking on prune sum}
To better understand the workings of the sparsity-locking scheme we investigated the evolution of the \textit{prune sum}, which is the sum of the absolute values of the weights pruned. For brevity, we present the evolution of prune sum for only \textit{ViT-B\_16-224}.

\begin{figure}[H]
    \centering
    \includegraphics[width=0.75\linewidth]{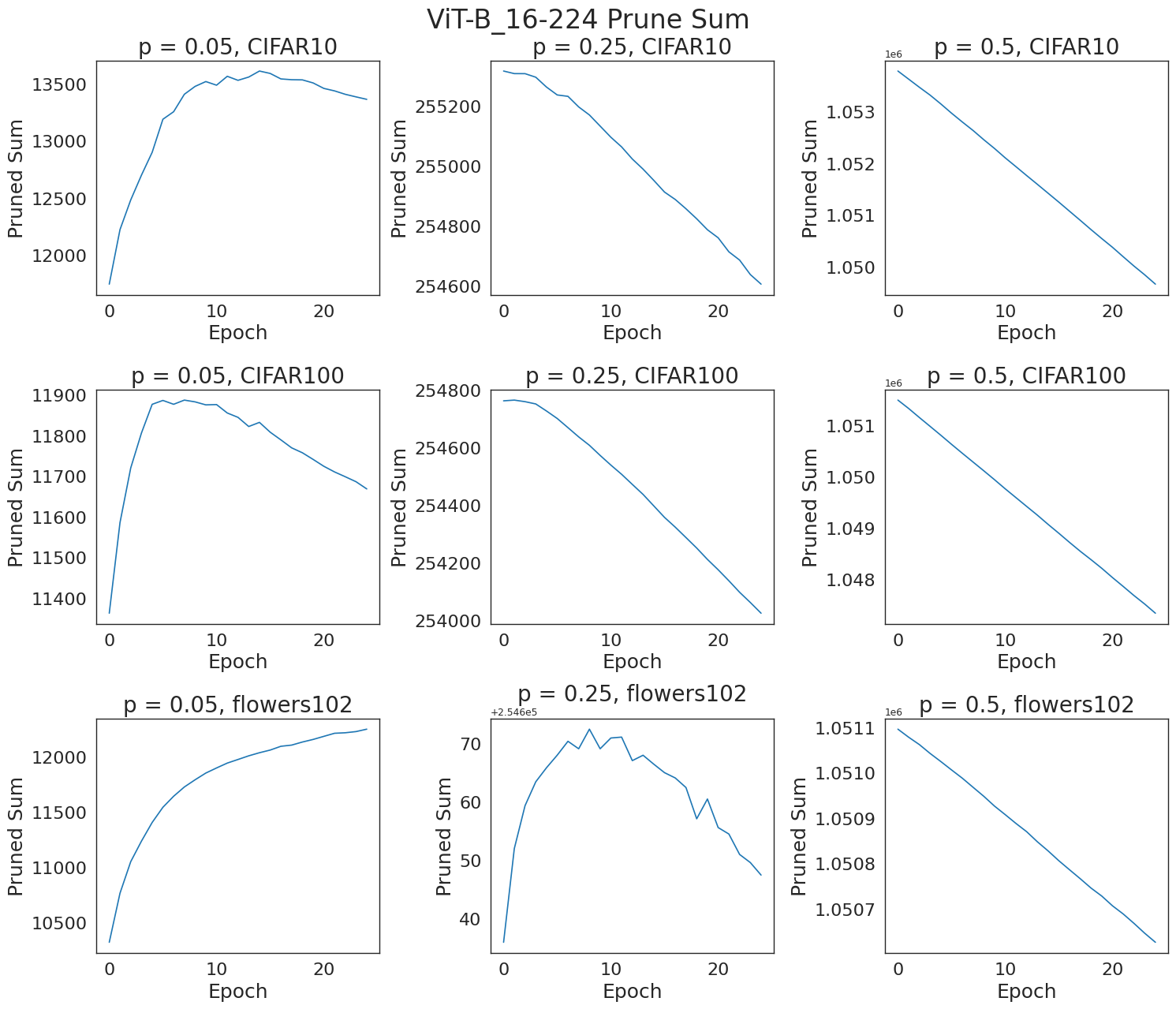}
    \caption{Evolution of Prune Sum during Sparsity-Locking (\textit{ViT-B\_16-224})}
    \label{fig:Pruning_Abs_Sum}
\end{figure}

Note that the $y$-axis scales of the different settings are vastly different. Interestingly, this diagnostic illuminates that there are two regimes for the sparsity-aware locking scheme, dependent on the value of $p$. For locking low-levels of sparsity and $p$, the locking procedure increases the absolute magnitudes of the $p$-smallest parameters which, to a first-order, increases their \textit{importance} in model inference. At larger values of $p$, increases to the prune sum are negligible, if at all, suggesting the locking procedure does not rely on the increasing of magnitudes.

\subsection{Parameter Transform Distributions}\label{sec:transform-distributions}

Distributions of the parameters for various parameter transformation methods, including view of significands and exponents for floating point numbers. "Detransformed" is detransformed with the same fingerprint that was transformed with, while "incorrectly detransformed" is detransformed with any other fingerprint. The Incorrectly Detransformed parameter distributions should look as close as possible to the Detransformed pararameter distributions for indistinguishability to hold. We evaluate that indistinguishability holds for the shuffle and directly estimated pre-transformed AES methods, weakly holds for Gaussian-assumed pretransformed AES, and does not hold for other methods.

\begin{figure}[H]
	\centering
	\adjustbox{width=\linewidth}{
		\input{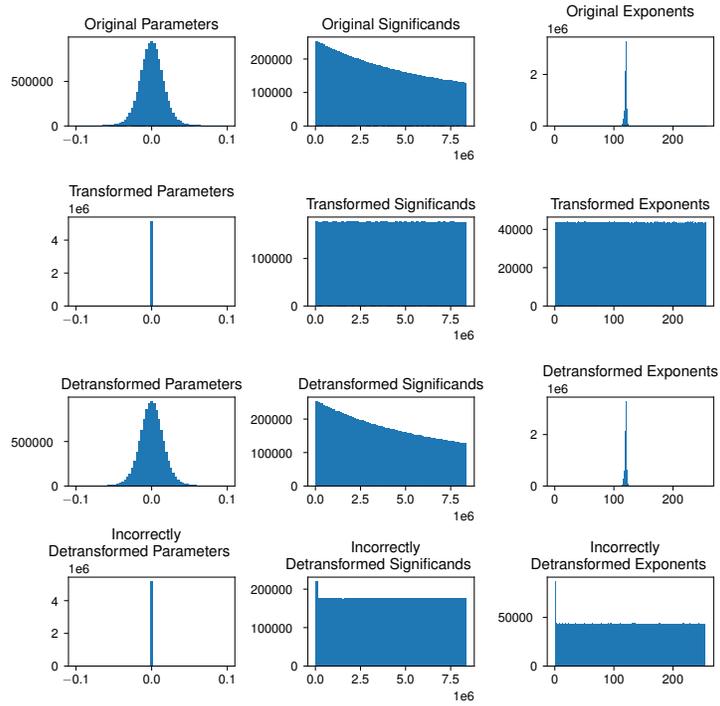}
	}
	\caption{AES encryption transformation method}
	\label{fig:transform-distributions-encrypt}
\end{figure}

\begin{figure}[H]
	\centering
	\adjustbox{width=\linewidth}{
		\input{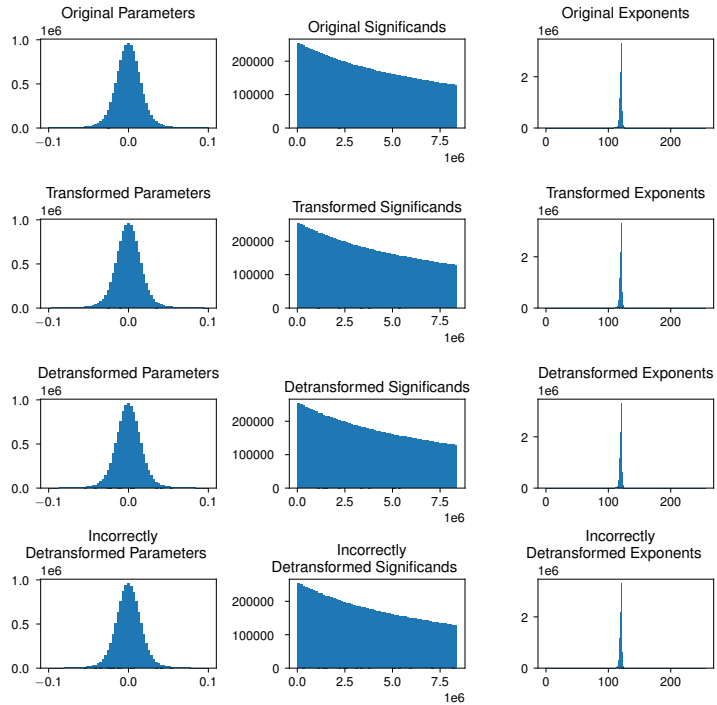}
	}
	\caption{Shuffle transformation method}
	\label{fig:transform-distributions-shuffle}
\end{figure}

\begin{figure}[H]
	\centering
	\adjustbox{width=\linewidth}{
		\input{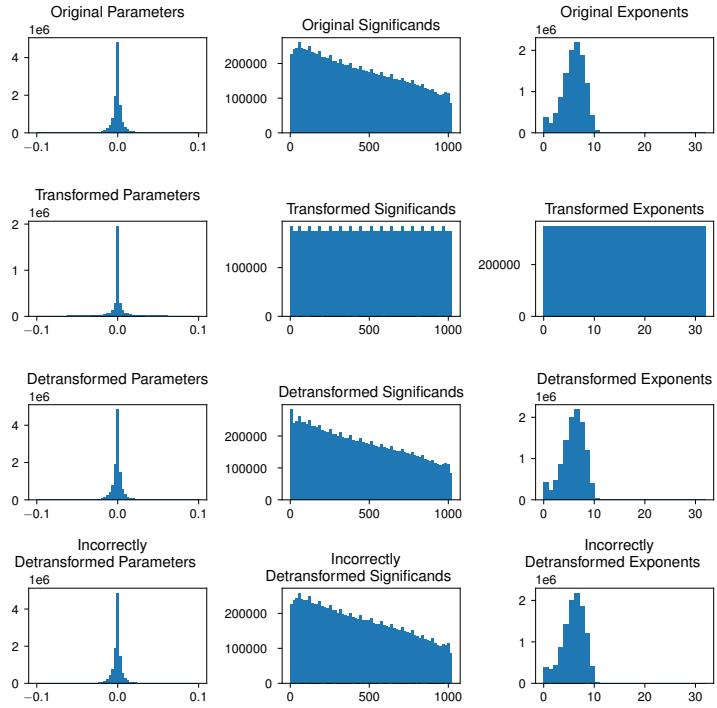}
	}
	\caption{Pretransformed AES encryption transformation method, with direct estimation of distribution, on 16-bit floating point}
	\label{fig:transform-distributions-pretransformed-direct-encrypt}
\end{figure}

\begin{figure}[H]
	\centering
	\adjustbox{width=\linewidth}{
		\input{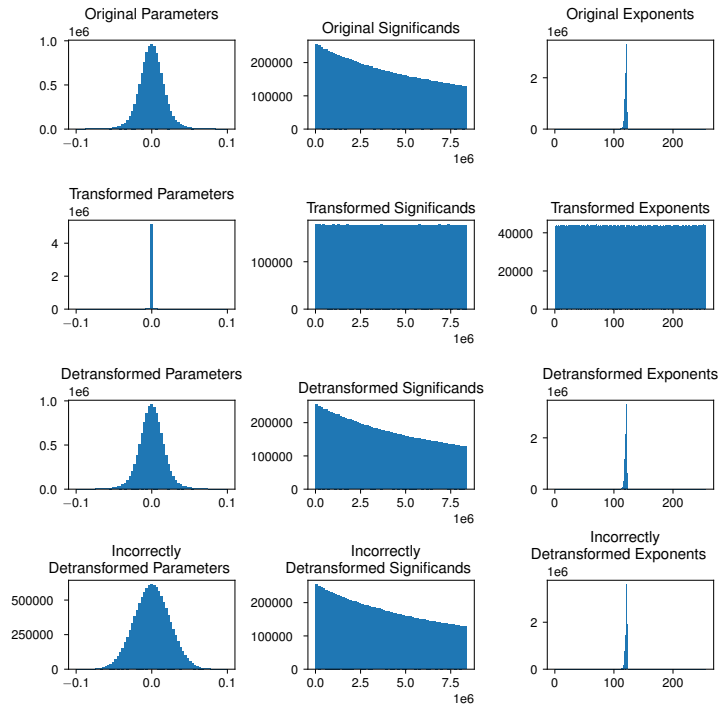}
	}
	\caption{Pretransformed AES encryption transformation method, assuming distribution to be Gaussian, on 32-bit floating point}
	\label{fig:transform-distributions-pretransformed-gaussian-encrypt}
\end{figure}

\newpage
\subsection{Finite Precision Fingerprints}\label{sec:finite-precision-fingerprint-data}

Fingerprints generated by code in~\Cref{sec:finite_precision_code}, on a number
of different devices with varying CUDA and PyTorch versions.

\begin{table}[H]
\centering
\caption{Finite precision fingerprint on various devices}
\adjustbox{width=\linewidth}{
\begin{tabular}{lll}
\toprule
	\textbf{Fingerprint (truncated)} & \textbf{Devices} \\
	\midrule

142c1a1a91c93ba1d6c02227e93a9905  & RTX 4080+torch1.13+cuda11.7 \\
3979ff885b639d070141ed3a829e49fd  & RTX 4000 Ada+torch1.13+cuda11.7 \\
~                                 & RTX A4000+torch1.13+cuda11.7 \\
405c153d574e01011c9f01c299af09b5  & RTX 3070+torch1.13+cuda11.7 \\
414168b2e9f45b44aeaf47a6d2f2d43f  & RTX 4000 Ada+torch2.2.0+cuda12.1 \\
467b11e341dd95e42f6386a6b6a2c9f9  & RTX 4070Ti+torch2.2.0+cuda12.1 \\
55265fbc99048cf94f3cecabf3209d73  & RTX 5000 Ada+torch2.2.0+cuda12.1 \\
58aebc551a1a3a4158b1063257bbb1d6  & RTX 3080+torch2.2.0+cuda12.1 \\
59342d03c054efbda09b1c7d1669ca7e  & RTX A4500+torch1.13+cuda11.7 \\
5bcb85507944291de0716edaa3e77c6b  & A40+torch1.13+cuda11.7 \\
~                                 & RTX A6000+torch2.0+cuda11.8 \\
~                                 & RTX A6000+torch1.13+cuda11.7 \\
~                                 & RTX 3090Ti+torch1.13+cuda11.7 \\
5d5b8d9d6897b31a99e40199404bdf4e  & RTX 6000 Ada+torch1.13+cuda11.7 \\
~                                 & NVIDIA L40S+torch1.13+cuda11.7 \\
68085f05ecc499655bf7923c8a7f65a2  & A40+torch2.2.0+cuda12.1 \\
~                                 & RTX A6000+torch2.2.1+cuda12.1 \\
~                                 & RTX A6000+torch2.2.0+cuda12.1 \\
~                                 & RTX A6000+torch2.1+cuda12.1 \\
~                                 & RTX A6000+torch2.1+cuda11.8 \\
~                                 & RTX 3090Ti+torch2.2.0+cuda12.1 \\
6f3b170b0ca899c52ac99df1eff2153f  & RTX A4500+torch2.2.0+cuda12.1 \\
812929c0b581eb634a88d7243bb4d442  & RTX 4090+torch1.13+cuda11.7 \\
834a709ba2534ebe3ee1397fd4f7bd28  & H100 NVL+torch1.13+cuda11.7 \\
8399893d7103588010da58597516475a  & H100 NVL+torch2.2.0+cuda12.1 \\
9e85db98fbf34dfad6d3712afe3f0aa2  & RTX 3090+torch1.13+cuda11.7 \\
b7f216f032ec3e598c436df9d51ed9c0  & RTX 4080+torch2.2.0+cuda12.1 \\
bb514adf79e70e5df9d839e3e16d9925  & RTX 6000 Ada+torch2.2.0+cuda12.1 \\
~                                 & NVIDIA L40S+torch2.2.0+cuda12.1 \\
bcc46fc1d5c8f3b912c75392e0221161  & RTX 4070Ti+torch1.13+cuda11.7 \\
bd3763175ad454cfe98684be54f952fd  & RTX 5000 Ada+torch1.13+cuda11.7 \\
ce51f90e8681521dac09984cac0cfd27  & RTX A5000+torch2.2.0+cuda12.1 \\
d45128ec2486355b452b7b8fc27453d9  & RTX A5000+torch1.13+cuda11.7 \\
d75f981870c15c5fd30fdc908f2c6fb2  & RTX 3070+torch2.2.0+cuda12.1 \\
d8392b1aa5cf26dfae05020d34f0073f  & RTX 4090+torch2.2.0+cuda12.1 \\
e6f5c511cd665dec0755c2cac6db3056  & RTX 3080+torch1.13+cuda11.7 \\
ef2135a16f9da8487fd69ac322d7053a  & RTX A4000+torch2.2.0+cuda12.1 \\
f51e6d276b5c634f18786d58ae8e5cd9  & RTX A6000+torch1.9+cuda11.1 \\
	\bottomrule
\end{tabular}
}

\label{tab:finite-precision-fingerprint}
\end{table}

\newpage
\subsection{Further Clock Fingerprints}\label{sec:clock-fingerprint-data}

More fingerprints generated by code in~\Cref{sec:clock_fingerprint_code}, on a
number of different devices with varying CUDA versions.

\begin{table}[H]
\centering
\caption{Further clock fingerprint tests}
\begin{tabular}{ll}
\toprule
	\textbf{Fingerprint}      & \textbf{Devices} \\
	\midrule
		4b85a  & RTX 3080+torch2.2.0+cuda12.1 \\
		~      & RTX A6000+torch2.2.1+cuda12.1 \\
		~      & RTX A6000+torch2.2.0+cuda12.1 \\
		~      & RTX A6000+torch2.1+cuda12.1 \\
		~      & RTX 3090Ti+torch2.2.0+cuda12.1 \\
		~      & A40+torch2.2.0+cuda12.1 \\
		~      & RTX A6000+torch1.9+cuda11.1 \\
		~      & RTX 4090+torch2.2.0+cuda12.1 \\
		4c85e  & RTX 4000 Ada+torch2.2.0+cuda12.1 \\
		~      & RTX 4070Ti+torch2.2.0+cuda12.1 \\
		4787c  & H100 NVL+torch2.2.0+cuda12.1 \\
		3f67c  & H100 NVL+torch1.13+cuda11.7 \\
		4be5c  & RTX A4500+torch2.2.0+cuda12.1 \\
		~      & RTX 3070+torch2.2.0+cuda12.1 \\
		~      & RTX A4000+torch2.2.0+cuda12.1 \\
		~      & RTX A5000+torch2.2.0+cuda12.1 \\
		44e5c  & RTX 4090+torch1.13+cuda11.7 \\
		~      & NVIDIA L40S+torch1.13+cuda11.7 \\
		~      & RTX 4080+torch1.13+cuda11.7 \\
		~      & RTX 6000 Ada+torch1.13+cuda11.7 \\
		~      & RTX 5000 Ada+torch1.13+cuda11.7 \\
		4545e  & RTX 4000 Ada+torch1.13+cuda11.7 \\
		~      & RTX 4070Ti+torch1.13+cuda11.7 \\
		4485c  & RTX A4500+torch1.13+cuda11.7 \\
		~      & RTX 3070+torch1.13+cuda11.7 \\
		~      & RTX A4000+torch1.13+cuda11.7 \\
		~      & RTX A5000+torch1.13+cuda11.7 \\
		4465a  & RTX 3080+torch1.13+cuda11.7 \\
		~      & RTX 3090Ti+torch1.13+cuda11.7 \\
		~      & A40+torch1.13+cuda11.7 \\
		~      & RTX 3090+torch1.13+cuda11.7 \\
		~      & RTX A6000+torch2.0+cuda11.8 \\
		~      & RTX A6000+torch1.13+cuda11.7 \\
		~      & RTX A6000+torch2.1+cuda11.8 \\
		4c25c  & RTX 4090+torch2.2.0+cuda12.1 \\
		~      & NVIDIA L40S+torch2.2.0+cuda12.1 \\
		~      & RTX 4080+torch2.2.0+cuda12.1 \\
		~      & RTX 6000 Ada+torch2.2.0+cuda12.1 \\
		~      & RTX 5000 Ada+torch2.2.0+cuda12.1 \\
 \bottomrule
\end{tabular}

\label{tab:clock-fingerprints-extra}
\end{table}

\newpage
\subsection{CUDA Code to Produce Clock Fingerprint}\label{sec:clock_fingerprint_code}

\begin{lstlisting}[language=C]
/* Copyright (c) 2024, Eleanor Clifford
 * Copyright (c) 2022, NVIDIA CORPORATION. All rights reserved.
 *
 * Redistribution and use in source and binary forms, with or without
 * modification, are permitted provided that the following conditions
 * are met:
 *    * Redistributions of source code must retain the above copyright
 *      notice, this list of conditions and the following disclaimer.
 *    * Redistributions in binary form must reproduce the above
 *      copyright notice, this list of conditions and the following
 *      disclaimer in the documentation and/or other materials
        provided with the distribution.
 *    * Neither the name of NVIDIA CORPORATION nor the names of its
 *      contributors may be used to endorse or promote products
 *      derived from this software without specific prior written
 *      permission.
 *
 * THIS SOFTWARE IS PROVIDED BY THE COPYRIGHT HOLDERS ``AS IS'' AND
 * ANY EXPRESS OR IMPLIED WARRANTIES, INCLUDING, BUT NOT LIMITED TO,
 * THE IMPLIED WARRANTIES OF MERCHANTABILITY AND FITNESS FOR A
 * PARTICULAR PURPOSE ARE DISCLAIMED.    IN NO EVENT SHALL THE
 * COPYRIGHT OWNER OR CONTRIBUTORS BE LIABLE FOR ANY DIRECT, INDIRECT,
 * INCIDENTAL, SPECIAL, EXEMPLARY, OR CONSEQUENTIAL DAMAGES
 * (INCLUDING, BUT NOT LIMITED TO, PROCUREMENT OF SUBSTITUTE GOODS OR
 * SERVICES; LOSS OF USE, DATA, OR PROFITS; OR BUSINESS INTERRUPTION)
 * HOWEVER CAUSED AND ON ANY THEORY OF LIABILITY, WHETHER IN CONTRACT,
 * STRICT LIABILITY, OR TORT (INCLUDING NEGLIGENCE OR OTHERWISE)
 * ARISING IN ANY WAY OUT OF THE USE OF THIS SOFTWARE, EVEN IF ADVISED
 * OF THE POSSIBILITY OF SUCH DAMAGE.
 */

// System includes
#include <assert.h>
#include <stdint.h>
#include <stdio.h>
#include <ieee754.h>
#include <limits.h>

// CUDA runtime
#include <cuda_runtime.h>

// helper functions and utilities to work with CUDA
#include <helper_cuda.h>
#include <helper_functions.h>

#define INNER_LOOP 512
#define NUM_BLOCKS 128
#define NUM_THREADS 8
#define NUM_TRIES 16

// This kernel does something which doesn't matter.
// The timing results are stored in device memory.
__global__ static void timedFunction(
  const float *input, float *output,
  clock_t *timer
) {
  extern __shared__ float shared[];

  const int tid = threadIdx.x;
  const int bid = blockIdx.x;

  // Copy input.
  shared[tid] = input[tid];
  shared[tid + blockDim.x] = input[tid + blockDim.x];

  if (tid == 0) timer[bid] = clock();

  // Do some stuff
  for (size_t i = 0; i < INNER_LOOP; i++) {
    for (int d = blockDim.x; d > 0; d /= 2) {
      __syncthreads();

      if (tid < d) {
        float f0 = shared[tid];
        float f1 = shared[tid + d];

        if (f1 < f0) {
        shared[tid] = (f0 + f1);
        }
      }
    }
  }

  // Write result.
  if (tid == 0) output[bid] = shared[0];
  __syncthreads();
  if (tid == 0) timer[bid + gridDim.x] = clock();

}

// Start the main CUDA Sample here
int main(int argc, char **argv) {
  cudaSetDevice(2);

  float *dinput = NULL;
  float *doutput = NULL;
  clock_t *dtimer = NULL;

  clock_t timer[NUM_BLOCKS * 2];
  float input[NUM_THREADS * 2];

  for (int i = 0; i < NUM_THREADS * 2; i++) {
    input[i] = (float)i;
  }

  long fastestClock = LONG_MAX;
  for (int j = 0; j < NUM_TRIES; j++) {

    checkCudaErrors(
        cudaMalloc((void **)&dinput, sizeof(float) * NUM_THREADS * 2)
    );
    checkCudaErrors(
        cudaMalloc((void **)&doutput, sizeof(float) * NUM_BLOCKS)
    );
    checkCudaErrors(
        cudaMalloc((void **)&dtimer, sizeof(clock_t) * NUM_BLOCKS * 2)
    );

    checkCudaErrors(
        cudaMemcpy(dinput, input, sizeof(float) * NUM_THREADS * 2,
                   cudaMemcpyHostToDevice)
    );

    timedFunction<<<
        NUM_BLOCKS, NUM_THREADS, sizeof(float) * 2 * NUM_THREADS
    >>>(dinput, doutput, dtimer);

    checkCudaErrors(
        cudaMemcpy(timer, dtimer, sizeof(clock_t) * NUM_BLOCKS * 2,
                   cudaMemcpyDeviceToHost)
    );

    checkCudaErrors(cudaFree(dinput));
    checkCudaErrors(cudaFree(doutput));
    checkCudaErrors(cudaFree(dtimer));

    for (int i = 0; i < NUM_BLOCKS; i++) {
      long t = (timer[i + NUM_BLOCKS] - timer[i]);
      if (t < fastestClock) {
        fastestClock = t;
      }
    }
  }

  printf("%x\n", (unsigned int)fastestClock);

  return EXIT_SUCCESS;
}
\end{lstlisting}

\subsection{Python Code to Produce Finite Precision Fingerprint}\label{sec:finite_precision_code}

\begin{lstlisting}[language=Python]
# Copyright (c) 2024, Eleanor Clifford
#                 and Ilia Shumailov
# MIT License

import torch
import hashlib
import numpy as np

dev = "cuda"

torch.manual_seed(0)
inp = torch.randn(1, 50, 50, 100)
m = torch.nn.Sequential(
  torch.nn.Linear(100, 1000),
  torch.nn.Linear(1000, 10000),
  torch.nn.Linear(10000, 10),
)

m = m.to(dev)
inp = inp.to(dev)

out = []

orig = None
with torch.no_grad():
  for i in range(1, 10):
    _input = torch.cat(i * [inp]).clone()
    output = m(_input)

    if orig is None:
      orig = output[0].clone().detach()

    out.append(float((orig - output).sum()))

print(hashlib.sha256(np.array(out)).hexdigest())
\end{lstlisting}

\newpage

\end{document}